%% file: manuscript.tex
\documentclass[%
 reprint,
 superscriptaddress,
 amsmath,amssymb,
 aps,longbibliography
]{revtex4-1}

\usepackage{graphicx}% Include figure files
\usepackage{dcolumn}% Align table columns on decimal point
\usepackage{physics}
\usepackage{float}
\usepackage{hhline}
\begin{document}

\title{Structural entropy and spatial decay of quasimodes in Vogel spirals}

% Marcus
\author{M. Prado}
\affiliation{Instituto de F\'isica, Universidade Federal do Rio de Janeiro, Rio de Janeiro-RJ, 21941-972, Brazil}
% Fabrizio
\author{F. Sgrignuoli}
\affiliation{Department of Electrical and Computer Engineering, Boston University, 8 Saint Mary\textsc{\char13}s Street, Boston, Massachusetts 02215, USA}
% Yuyao
\author{Y. Chen}
\affiliation{Department of Electrical and Computer Engineering, Boston University, 8 Saint Mary\textsc{\char13}s Street, Boston, Massachusetts 02215, USA}
% Luca
\author{L. Dal Negro}
\email{dalnegro@bu.edu}
\affiliation{Department of Electrical and Computer Engineering, Boston University, 8 Saint Mary\textsc{\char13}s Street, Boston, Massachusetts 02215, USA}
\affiliation{Division of Material Science and Engineering, Boston University, 15 Saint Mary\textsc{\char13}s Street, Brookline, Massachusetts 02446, USA}
\affiliation{Department of Physics, Boston University, 590 Commonwealth Avenue, Boston, Massachusetts 02215, USA}
% Felipe
\author{F. A. Pinheiro}
\email{fpinheiro@if.ufrj.br}
\affiliation{Instituto de F\'isica, Universidade Federal do Rio de Janeiro, Rio de Janeiro-RJ, 21941-972, Brazil}

%%%%%%%%%%%%%%%%%%%%%%%%%%%%%%%%%%%%%%%%%%%%%%%%%%%%%%%%%%
%%%%%%%%%%%%%%%%%%		      Abstract             %%%%%%%%%%%%%%%%%%%%
%%%%%%%%%%%%%%%%%%%%%%%%%%%%%%%%%%%%%%%%%%%%%%%%%%%%%%%%%%
\begin{abstract}
We investigate the spatial decay and temporal localization properties of quasimodes (i.e., scattering resonances) of two-dimensional Vogel spirals, composed of deterministic, aperiodic arrays of electric dipoles. By determining the structural entropy and localization maps of Vogel spirals using the Green's matrix method, we show that three distinctive decay types of quasimodes coexist in Vogel spirals: exponential, power-law, and Gaussian. While the exponential and the power-law decays typically occur in disordered media and multifractal systems, respectively, the Gaussian decay is demonstrated to characterize, on average, the most localized quasimodes of Vogel spirals, both spatially (smallest participation ratios) and temporarily (longest lifetimes). These decay forms are demonstrated by a no-fitting analysis of the localization maps, independently corroborated by calculating the electric field in real space, which also provides a direct evidence of the algebraic spatial decay of critical quasimodes.  Altogether our findings unveil a rich spectrum of both long-lived and spatially localized quasimodes that coexist in Vogel spirals and can be of direct relevance to novel optical functionalities for applications to light sources and sensing devices.
\end{abstract}

\maketitle

%%%%%%%%%%%%%%%%%%%%%%%%%%%%%%%%%%%%%%%%%%%%%%%%%%%%%%%%%%
%%%%%%%%%%%%%%%%%%		      Introduction             %%%%%%%%%%%%%%%%%%%%
%%%%%%%%%%%%%%%%%%%%%%%%%%%%%%%%%%%%%%%%%%%%%%%%%%%%%%%%%%
\section{Introduction}
The successful design of photonic nanostructures crucially depends on the efficient characterization of the electromagnetic modes they can support. On one hand, ordered optical metamaterials with periodically arranged units of sub-wavelength dimensions have been demonstrated to exhibit unusual optical functionalities, with many applications~\cite{zheludev2012metamaterials}. On the other hand, disordered optical media exhibit fascinating analogies between quantum and classical transport wave phenomena, such as Anderson localization~\cite{segev2013anderson,lagendijk2009fifty}, with applications such as random lasers~\cite{wiersma2008physics}. However, disordered optical structures lack the reproducible and predictable behavior that is necessary for designing many advanced optical devices.

Deterministic aperiodic systems, artificial optical media with a tunable degree of aperiodic structural order generated by deterministic mathematical rules, have emerged as an alternative
material platform for designing photonic devices~\cite{Gopinath,Lee,noh2011lasing,DalNegroReview,DalNegroBook,MaciaBook,MaciaBook2,DalNegroCrystals,sgrignuoli2020multifractality,sgrignuoli2020subdiffusive,TPSEjosab}. Indeed, these structures present unique optical properties that do not occur in either periodic or disordered systems, such as fractal transmission spectra and anomalous transport properties, which include multifractality of light~\cite{sgrignuoli2020multifractality}, subdiffusive transport~\cite{sgrignuoli2020subdiffusive}, and light localization~\cite{sgrignuoli2019localization}. As a result of their unique functionalities, deterministic aperiodic systems have been employed in many applications such as lasing~\cite{vardeny2013optics}, optical sensing~\cite{razi2019optimization,Lee,Gopinath}, photo-detection~\cite{trevino2011circularly}, and optical imaging~\cite{huang2007optical}.

Among several candidates of deterministic aperiodic optical systems, the robust platforms of Vogel spiral arrays stands out for its versatility and the possibility to tailor its structural order between short-range correlated amorphous systems and uncorrelated random ones~\cite{trevino2011circularly,dal2012analytical,lawrence2012control,liew2011localized,christofi2016probing,Pollard}. Vogel spirals are deterministic structures with Fourier spectra characterized by diffuse circular rings and do not have well-defined Bragg peaks, in contrast to photonic crystals and quasicrystals~\cite{DalNegroCrystals}. The positions of particles in Vogel spiral arrays are generated in polar coordinates according to $r_{n}=a_{0}\sqrt{n}$, $\theta_{n}=n \alpha$, where $n=1,2,\ldots$ is an integer, $a_{0}$ is the scaling factor determining particle separation, and $\alpha$ is the divergence angle~\cite{Naylor,sgrignuoli2019localization,MaciaBook,dal2012analytical,lawrence2012control}.  This angle specifies the constant aperture between successive point scatterers in the array. Since $\alpha$ is irrational, Vogel spiral point patterns
lack both translational and rotational symmetry. Any irrational number ($\xi$) can be used to produce the divergence angle ($\alpha^{\circ}$, in degrees) according to the relationship $\alpha^{\circ}=360^{\circ}-\textbf{frac}(\xi)\cdot{360}^{\circ}$
where $\textbf{frac}(\xi)$ denotes the fractional part of $\xi$. The value of $\xi$ defines the four types of structures considered here, which exhibit different degrees of structural order:  $\xi=(1+\sqrt{5})/2$ (golden mean), $\xi=(2+\sqrt{8})/2$, $\xi=\pi$, and  $\xi=(5+\sqrt{29})/2$, for the GA (Golden Angle)-, $\tau$-, $\pi$- and $\mu$-spirals respectively~\cite{DalNegroCrystals}. 
Vogel spirals support distinctive optical resonances, with a broad range of lifetimes and spatial profiles, such as  critical modes, which are expected to show algebraic, power-law envelope decay and multi-fractal field intensity oscillations~\cite{noh2011lasing,Ryu1992,Desideri1989,Macia1999,mahler2010quasi}. Localized quasimodes also occur in Vogel spirals, leading to a light localization  transition~\cite{sgrignuoli2019localization}.

Despite the relevance and applicability of Vogel spirals in optics of aperiodic media, the full characterization of temporal localization and spatial decay of their electromagnetic modes has never been achieved so far.  In the present work we address this important issue using an alternative approach to probe the spatial decay of the quasimodes (i.e., scattering resonances), based on the so-called localization maps~\cite{Pipek1992}. Localization maps are based on the relation between the structural entropy and the participation ratio of modes, which allows for a general (independent of the system size and geometry) and no-fitting method to characterize the spatial decay of eigenstates of an arbitrary lattice, regardless of its dimensionality~\cite{Pipek1992}. This method has been successfully applied to characterize localization in quantum systems~\cite{santos2010localization}, metal-insulator transitions~\cite{aulbach2004phase,varga1995shape}, and electronic wave functions in quasiperiodic lattices~\cite{rieth1998numerical}. By means of localization maps of quasimodes of Vogel spirals in two dimensions, which have been calculated using the Green's matrix method~\cite{rusek2000random,pinheiro2004probing,pinheiro2008statistics,goetschy2011non,skipetrov2011eigenvalue,DalNegroCrystals,maximo2015spatial}, we identify three distinctive forms of spatial decay: Gaussian, power-law, and exponential, which so far has been usually associated with disordered, Anderson localized systems. These types of decays are confirmed by independent calculations of the electric field in real space. By correlating the spatial decay of the quasimodes with their spectral overlap, which is measured by computing the ratio between its decay rate to the mean level spacing, we conclude that in Vogel spirals the most localized quasimodes, both temporarily (i.e., modes with longest lifetimes) and spatially (i.e., modes with smallest participation ratios), exhibit Gaussian decay on average. 

%%%%%%%%%%%%%%%%%%%%%%%%%%%%%%%%%%%%%%%%%%%%%%%%%%%%%%%%%%
%%%%%%%%%%%%%%%%%%		      Methodology             %%%%%%%%%%%%%%%%%%%%
%%%%%%%%%%%%%%%%%%%%%%%%%%%%%%%%%%%%%%%%%%%%%%%%%%%%%%%%%%
\section{Methodology}
%%%%%%%%%%%%%%%%%%%%%%%%%%%%%%%%%%%%%%%%%%%%%%%%%%%%%%%%%%
%%%%%%%%%%%%%%%%%%		       Green's matrix method in 2D arrays             %%%%%%%%%%%%%%%%%%%%
%%%%%%%%%%%%%%%%%%%%%%%%%%%%%%%%%%%%%%%%%%%%%%%%%%%%%%%%%%
Planar Vogel spiral arrays in which light scattering and propagation are confined to two dimensions may be actually realized in a number of systems such as arrays of dielectric nanocylinders~\cite{Lawrence2012pillars}, planar arrangements of scatterers in microwave cavities (e.g., \cite{aubry2020experimental}), photonic crystals (e.g., \cite{busch2007periodic}), and laser-cooled atoms in optical cavities~\cite{maximo2015spatial}. In Sec.~\ref{greensec} we present the Green's matrix method in two-dimensions, which is a general approach that is used to model electromagnetic propagation problems. In Sec.~\ref{Mode2DGMT} we specifically consider light propagation in deterministic Vogel spiral arrays composed of dielectric cylinders within the framework of the generalized Mie theory, i.e., 2D-GMT~\cite{AsatryanPRE}. Using these two approaches, Green's matrix method and 2D-GMT, one can independently determine the electromagnetic quasimodes of the system under study. Then, one can investigate their spatial decay using the localization map method, which will be presented in Sec.~\ref{locmapssec}. Localization maps analysis using both methods will be presented, discussed, and compared in Sec.~\ref{ResulsDDiscussion}.

\subsection{Green's matrix method in 2D arrays}\label{greensec}
The Green's matrix method allows one to determine the scattering resonances of an arbitrary system of $N$ point scatterers~\cite{rusek2000random,RusekPRE2D,skipetrov2011eigenvalue,SkipetrovPRL,pinheiro2004probing}. Within this model, information about the scattering resonances of the system can be extracted from the spectrum of the Green’s matrix. Multiple scattering is treated exactly and the only approximation is to consider the scatterers as electric point dipoles. In two-dimensions the Green's matrix reads~\cite{RusekPRE2D,maximo2015spatial}
%%%%%%%%%%%%%%%%%%%%%% Green Matrix %%%%%%%%%%%%%%%%%%%%%% 
\begin{equation}\label{Green}
G_{ij}= i\delta_{ij} + i(1-\delta_{ij}) H_0(k_0|\textbf{r}_i-\textbf{r}_j|),
\end{equation}
%%%%%%%%%%%%%%%%%%%%%%%%%%%%%%%%%%%%%%%%%%%%%%%%%%%
where its off-diagonal elements describe wave propagation in free space between two point dipoles. In equation (\ref{Green}), the symbol $\delta_{ij}$ is the Kronecker delta function, $H_0(k_0|\textbf{r}_i-\textbf{r}_j|)$ is the zero-order Hankel function of the first kind, $k_0$ the wavevector of light, and $\textbf{r}_i$ the position of the $i^{th}$-scatterer in the array. The Green's matrix model in two-dimensions describes the electromagnetic propagation and scattering properties of infinite arrays of dielectric cylinders when excited by TM-polarized waves \cite{RusekPRE2D,maximo2015spatial,Leseur}. Even though the 2D model defined by the matrix (\ref{Green}) does not take into account the vector nature of light \cite{sgrignuoli2019localization,sgrignuoli2020subdiffusive,christofi2016probing,SkipetrovPRL}, it still provides useful information on light localization in 2D disordered media \cite{RusekPRE2D}, transparency in high-density hyperuniform materials \cite{Leseur}, correctly describes the coupling between one or several two-level atoms in a structured reservoir \cite{Caze,Bouchet}, and allows the design of aperiodic arrays for the efficient generation of multi-frequency two-photon processes~\cite{TPSEjosab}. 

To investigate the temporal localization and spatial decay of the electromagnetic quasimodes of Vogel spirals, we have numerically diagonalized the $N\times N$ Green's matrix, Eq.\,(\ref{Green}). The real and imaginary part of its complex eigenvalues $\Lambda_n$ ($n\in$ 1, 2, $\cdots$, N) correspond to the relative frequencies $(\omega_0-\omega_n)/\Gamma_0$ and widths $\Gamma_n/\Gamma_0$ of the scattering resonances, respectively \cite{RusekPRE2D,Lagendijk}. Here, $\Gamma_0$ and $\omega_0$ are the resonant width and the resonant frequency of a single electric dipole. The eigenvectors $\psi_n$ are their scattering resonances.

%%%%%%%%%%%%%%%%%%%%%%%%%%%%%%%%%%%%%%%%%%%%%%%%%%%%%%%%%%
%%%%%%%%%%%%%%%%%%		       Calculation of electromagnetic quasimodes in real space              %%%%%%%%%%%%%%%%%%%%
%%%%%%%%%%%%%%%%%%%%%%%%%%%%%%%%%%%%%%%%%%%%%%%%%%%%%%%%%%
\subsection{Calculation of electromagnetic quasimodes in real space}\label{Mode2DGMT}
To evaluate the electromagnetic quasimodes supported by the Vogel spirals, we used an efficient algorithm based on the rigorous solution of the 2D scattering problem in the framework of the generalized Mie theory, i.e., 2D-GMT \cite{AsatryanPRE}. Within this approach, the electromagnetic field is expanded in terms of Bessel-Fourier multipolar basis functions in the coordinate system centered at the origin of each individual cylinders. Using the Graf's addition theorem for Bessel functions and imposing the field continuity conditions at the boundary of each cylinder, we can obtain the polarization-dependent Mie coefficients by solving the linear system \cite{Gagnon}:
\begin{equation}\label{Teq}
\textbf{T} \textbf{b}=\textbf{a}_0
\end{equation}
with $\textbf{a}_0=a_{n\ell}/J_\ell(k_0R_n)$ and  $\textbf{b}=b_{n\ell}/J_\ell(k_0R_n)$, where $a_{n\ell}$ and $b_{n\ell}$  are the source and scattering Mie coefficients of the $n^{th}$-cylinder. Moreover, $k_0$ is the plane wave incident wavenumber and $R_n$ denotes the radius of the $n^{th}$-cylinder. The function $J_\ell(x_{n})$ is the cylindrical Bessel function of the first kind of order $\ell$ and $x_{n}=k_0R_{n}$ denotes the size parameter of the $n^{th}$-cylinder. The matrix $\textbf{T}$ describes the material and the geometrical properties of the scattering medium and it is defined as~\cite{doicu2006light}:

\begin{equation}\label{Tmatrix}
\begin{split}
\textbf{T}_{nn^\prime}^{\ell\ell^\prime}=\;&\delta_{nn^\prime}\delta_{\ell\ell^\prime}-(1-\delta_{nn^\prime})e^{i(\ell^\prime-\ell)\phi_{nn^\prime}}H_{\ell-\ell^\prime}(k_0r_{nn^\prime})\\
& s_{n\ell}\frac{J_{\ell^\prime}(k_0R_{n^\prime})}{J_{\ell}(k_0R_{n})},
\end{split}
\end{equation}
where $H_{\ell-\ell^\prime}(k_0r_{nn^\prime})$ is the Hankel function of the first kind of order $\ell-\ell^\prime$, $r_{n n^\prime}$ is the center-to-center distance between the cylinders identified by the indexes $n$ and $n^\prime$, the function $\phi_{n n^\prime}$ is the angular position of the $n^\prime$ cylinder in the reference frame of the cylinder $n$, and the parameter $s_{n\ell}$ is equal to $-[J_\ell^\prime(k_0R_n)-\Gamma_{n\ell}J_\ell(k_oR_n)]/[H_\ell^\prime(k_0R_n)-\Gamma_{n\ell}H_\ell(k_0R_n)]$. Here, the prime superscript indicates differentiation with respect to the argument, $\Gamma_{n\ell}=[\xi_nk_nJ_\ell^\prime(k_nR_n)]/k_0J_\ell(k_nR_n)$, while $k_n$ indicates the wavenumber inside the $n^{th}$-nanocylinder and $\xi_n$ is a parameter that depends on the polarization of the incident light. Specifically, in the transverse magnetic case $\xi_n=\mu_0/\mu_n$, while $\xi_n=\epsilon_0/\epsilon_n$ when a transverse electric incident filed is considered. The symbols $\mu_0$ and $\mu_n$ ($\epsilon_0$ and $\epsilon_n$) are, respectively, the vacuum and the nanocylinders permeability (permittivity). The computational effort required to solve the matrix equation (\ref{Teq}) is proportional to the number of cylinders, their separation distances, and the maximum multipolar order at which the infinite series were truncated. The T-matrix expression reported in (\ref{Tmatrix}) ensures numerical stability, even when a large number of multipolar orders is required. In the present work, the T-matrix was calculated by truncating the multipolar expansion order to $\ell_{max}=1$ (i.e., the field summation runs from $-\ell_{max}$ to $\ell_{max}$ yielding $2\ell_{max}+1=3$ multipolar orders). This truncation ensures the validity of the dipole approximations as further discussed in Sec.~\ref{ResulsDDiscussion}. Notice that Eq.\,(\ref{Tmatrix}) is a matrix composed of $N \times N$ blocks, where $N$ is the total number of scattering elements. Each block has a dimension of $(2l_{max} + 1) \times (2l_{max} + 1)$. More mathematical details on the utilized method can be found in  Refs.~\cite{Gagnon,AsatryanPRE,Wriedt2012}, while information regarding its applications to nanophotonics can be found, for example, in Refs.~\cite{Gopinath,Lee,noh2011lasing,DalNegroReview,DalNegroBook,MaciaBook,MaciaBook2,razi2019optimization,Trojak1,Trojak2,ringler2008shaping,doicu2006light}.

Calculating the resonant electromagnetic quasimodes requires solving the homogeneous equation $\textbf{T} \textbf{b}=0$ \cite{Trojak1,Trojak2,TPSEjosab,Gagnon,AsatryanPRE}. Specifically, we have to find the complex $\tilde{k}$ values for which the relation $\det[\textbf{T}(\tilde{k})]=0$ is satisfied \cite{Gagnon}. The complex eigenvalue $\tilde{k}$ has a direct physical interpretation, namely its real part is related to the wavenumber of the mode, while its imaginary part is proportional to its spectral width $\Gamma=2c\abs{\Im[\tilde{k}]}$ ($c$ is the speed of light). Modes with imaginary parts closer to zero are the ones with the longest lifetimes and hence correspond to the highest quality factors, which is defined as $Q=\Re[\tilde{k}]/2\abs{\Im[\tilde{k}]}$ \cite{vogel2006quantum,Gagnon}.

%%%%%%%%%%%%%%%%%%%%%%%%%%%%%%%%%%%%%%%%%%%%%%%%%%%%%%%%%%
%%%%%%%%%%%%%%%%%%		       Structural entropy and localization maps              %%%%%%%%%%%%%%%%%%%%
%%%%%%%%%%%%%%%%%%%%%%%%%%%%%%%%%%%%%%%%%%%%%%%%%%%%%%%%%%
\subsection{Structural entropy and localization maps}\label{locmapssec}

Localization maps have been introduced in Ref.~\cite{Pipek1992} to probe the spatial decay and extent of eigenfunctions in disordered media. The latter property is quantified by the Mode Spatial Extent (MSE), also known as participation ratio~\cite{SgrignuoliACS,Mirlin}. The MSE of the $n^{th}$-(normalized) quasimode $\psi_n$ is defined as 
\begin{equation}
    MSE_n = \frac{\left[\sum_{i=1}^N \abs{\psi_n(i)}^2 \right]^2}{\sum_{i=1}^N \abs{\psi_n(i)}^4},
    \label{qn}
\end{equation}
where $N$ is the total number of sites of the system. The filling factor $q_n = MSE_n/N$ defines a quantity independent of the system size.

The other parameter needed for characterizing the spatial decay of the $n^{th}$-quasimode is the structural entropy entropy $S_{str_n}$, which is defined as~\cite{Pipek1992}
\begin{equation}
    S_{str_n} = -\sum_{i=1}^N \abs{\psi_n(i)}^2 \ln{\abs{\psi_n(i)}^2} - \ln{MSE_{n}}.   
\label{str} 
\end{equation}
Graphs of $S_{str_n}$ as a function of $q_{n}$, the so-called localization maps, allow one to predict the existence of modes with arbitrary types of spatial decay. There exists a non-trivial relation between these two quantities for a prescribed envelope function $f(r)$ that describes the decay of the quasimode, with $r$ being the distance to the maximum of intensity~\cite{Pipek1992}. In the localization map, modes with the same type of spatial decay should lie around the same reference curve, which can be analytically or numerically calculated. Therefore, this constitutes a systematic method of probing the average decay of quasimodes that do not rely on fitting the spatial profile of the electric field of every single mode.

In order to access information on temporal localization of each quasimode one should determine $g_n$, defined in \cite{Skipetrov2016} as
\begin{equation}
    g_n = \frac{\Gamma_n}{\overline{\Delta \omega}} 
    = \frac{\Im\Lambda_n}{\overline{\Re\Lambda_n-\Re\Lambda_{n-1}}},
    \label{conductance}
\end{equation}
where $\overline{\cdots}$ denotes a band average in the vicinity of $\omega_n$ and the eigenvalues are ordered such that $\Re\Lambda_n > \Re\Lambda_{n-1}$. $ g_n $ is essentially the inverse lifetime of the quasimode normalized by the mean level spacing.

%%%%%%%%%%%%%%%%%%%%%%%%%%%%%%%%%%%%%%%%%%%%%%%%%%%%%%%%%%
%%%%%%%%%%%%%%%%%%		       Results and Discussions              %%%%%%%%%%%%%%%%%%%%
%%%%%%%%%%%%%%%%%%%%%%%%%%%%%%%%%%%%%%%%%%%%%%%%%%%%%%%%%%
\section{Results and Discussions}\label{ResulsDDiscussion}

\begin{figure}[b!h!]
  \centering
    \includegraphics[width = \linewidth]{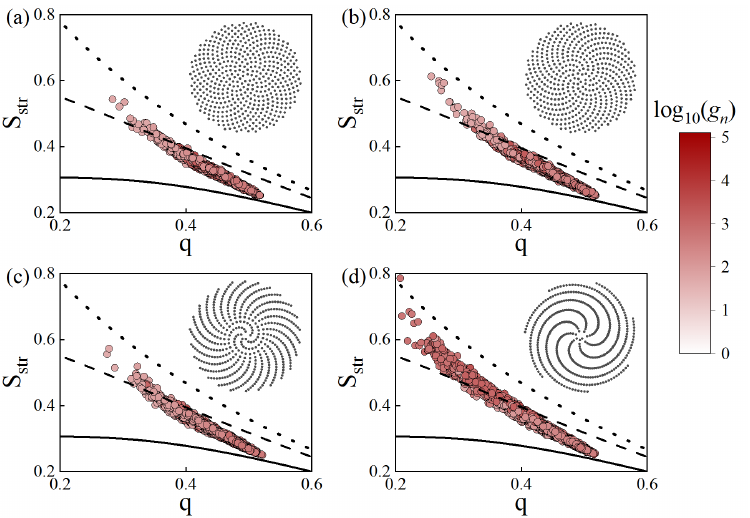}
    \caption{Structural entropy $S_{str}$ as a function of the filling factor $q$ for all the eigenvectors of the Green's matrix (\ref{Green}) for Vogel spiral arrays with 5000 point electric dipoles and optical density $\rho\lambda^2 = 0.1$. Panels (a)-(d) refer to the GA spiral, $\tau$ spiral, $\mu$ spiral, and $\pi$ spiral, respectively, which are depicted in the insets. The localization maps are color-coded according to the $\log_{10}$ values of $g_n$. The black lines correspond to the reference curves: Gaussian $f(r) = \exp(-r^{2})$ (solid line), exponential $f(r) = \exp (-r)$ (dashed line), and power-law $f(r) = (1+r)^{-3}$ (dotted line) in two dimensions.}
    \label{fig:locmap_lowOD}
\end{figure}

In Fig.~\ref{fig:locmap_lowOD} the localization maps for four types of Vogel spirals (Golden Angle, $\tau$, $\mu$, and $\pi$) are shown, where the structural entropy $S_{str}$ [Eq.\,(\ref{str})] is calculated as a function of the filling factor $q$  [Eq.\,(\ref{qn})] for optical density $\rho \lambda^{2}=0.1$, with $\rho$ the number density of dipoles and $\lambda$ the wavelength. We have verified that all localization maps are within the allowed domain of values $0 < q \leq 1$ and $0 \leq S_{str} \leq -ln(q)$~\cite{Pipek1992}. The color code indicates the value of $g_n$ of each quasimode, Eq.\,(\ref{conductance}).
Figure~\ref{fig:locmap_lowOD} reveals that the quasimodes' lifetimes are sufficiently low so that $g_n > 1$ for all investigated Vogel spirals. In disordered systems, $g_n$ is the Thouless conductance of each quasimode, and for $g_n > 1$ quasimodes strongly overlap spectrally and Anderson localization does not occur~\cite{Skipetrov2016}. The fact that $g_n > 1$ for the quasimodes of deterministic Vogel spirals suggests that localization within the array plane does not occur for this low optical density $\rho \lambda^{2}=0.1$, so that  waves rapidly leak through the system boundaries. From the localization maps it is not possible to identify a clear spatial decay of quasimodes, which are mostly extended and strongly overlapping in space.

\begin{figure}[b!h!]
    \centering
    \includegraphics[width = \linewidth]{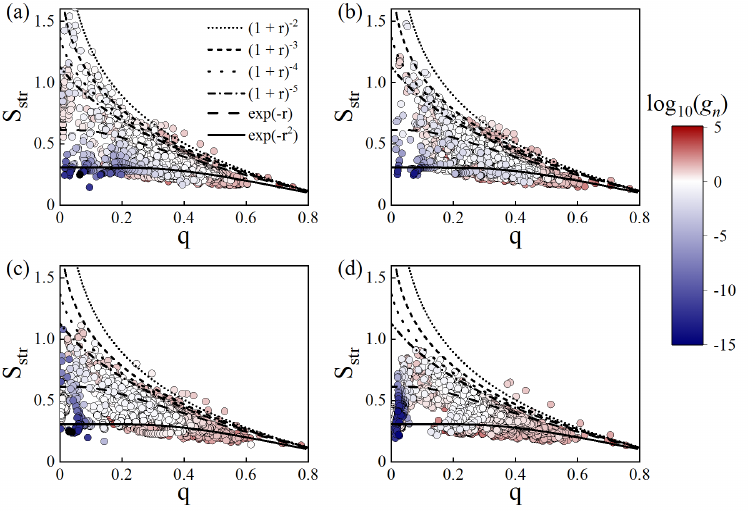}   
     \caption{Structural entropy $S_{str}$ as a function of the filling factor $q$ for all the eigenvectors of the Green's matrix (\ref{Green}) for Vogel spiral arrays with 5000 point electric dipoles and optical density $\rho\lambda^2 = 30$. Panels (a)-(d) refer to the GA spiral, $\tau$ spiral, $\mu$ spiral, and $\pi$ spiral, respectively. The localization maps are color-coded according to the $\log_{10}$ values of $g_n$. The black lines correspond to the reference curves: Gaussian $f(r) = \exp(-r^{2})$ (solid line), exponential $f(r) = \exp (-r)$ (long-dashed line), and power-law [$f(r) = (1+r)^{-2}$ (dotted line), $f(r) = (1+r)^{-3}$ (dashed line), $f(r) = (1+r)^{-4}$ (short-dashed line), $f(r) = (1+r)^{-5}$ (dash-dotted line)] in two dimensions.}
    \label{fig:locmap_highOD}
\end{figure}

This scenario is completely modified when one increases the optical density, as shown in Fig.~\ref{fig:locmap_highOD} where the localization maps for Vogel spirals are calculated for  $\rho \lambda^{2}=30$. In this case, the localization maps for the four Vogel spirals indicate the coexistence of three different types of spatially decaying quasimodes: Gaussian, power-law, and exponential. Indeed, localization maps in Fig.~\ref{fig:locmap_highOD} show that there exist many quasimodes that are well-described by the no-fitted, reference curves for Gaussian $f(r) = \exp(-r^{2})$, exponential $f(r) = \exp (-r)$, and power-law: $f(r) = (1+r)^{-\beta}$, where $\beta=2,3,4,5$. These functions characterize the decay shape of the envelope of the quasimodes~\cite{Pipek1992}. In this work, we show that exponentially localized modes may also exist in deterministic Vogel spirals, in much the same way they occur in disordered systems. However, in contrast to the uncorrelated case, these modes do not typically correspond to the ones with longest lifetimes, i.e., the most temporally localized. Rather we find that, for Vogel spirals, on average, quasimodes that exhibit Gaussian decay are not only the most temporally localized (for which $g_n <1$) but also the most spatially localized, corresponding to the smallest values of $q$ (smallest MSEs).  Localization maps also demonstrate that quasimodes with power-law decay do exist in Vogel spirals, with different degrees of spatial extent, i.e., corresponding to a broad range of values of $q$. These modes, also known as critical modes, are long-lived and extended modes that exhibit local fluctuations and spatial oscillations over multiple length scales~\cite{sgrignuoli2020multifractality,Ryu1992,Desideri1989,Macia1999}. The existence of power-law decaying quasimodes are expected to occur in aperiodic photonic media~\cite{DalNegroCrystals,TrevinoMF}, but a direct evidence of their existence in Vogel spirals has not been demonstrated so far. Altogether, the analysis of localization maps reveals that Vogel spirals exhibit a rich variety of localized quasimodes, with distinct spatial decays and different degrees of temporal localization.   

To establish a direct link with our previous results based on the Green's matrix method, we have selected the nanocylinder size, material, and averaged interparticle separation by evaluating the Purcell factor $P(\textbf{r}_p;\omega)=\rho(\textbf{r}_p;\omega)/\rho_0(\omega)$ 
with two complementary methods. Here,  $\rho(\textbf{r}_p;\omega)$ and $\rho_0(\omega)=\omega/2\pi c^2$ are the local density of optical states (LDOS) and the free-space local density of states, respectively. We have compared the $P(\textbf{r}_p;\omega)$ of 500 dielectric identical nanocylinders of relative permittivity $\varepsilon=10.5$ embedded in air \cite{TPSEjosab} to the Purcell enhancement of an assembly of 500 point electric dipoles characterized by the electric polarizability $\alpha(\omega)=-4\Gamma_0c^2/[\omega_0(\omega^2-\omega_0^2+i\Gamma_0\omega^2/\omega_0)]$ \cite{Leseur}. While in the 2D-GMT the LDOS calculation requires solving the linear system\,(\ref{Teq}) with the input coefficients $\textbf{a}_0$ describing a line source and then evaluating the total field at the position of the excitation $\textbf{r}_p$ \cite{TPSEjosab}, in the Green's matrix method the LDOS can be computed by solving the self-consistent equation:
\begin{equation}\label{GreenEscat}
E_i=\mu_0\omega^2G(\textbf{r}_i,\textbf{r}_p;\omega)p+\frac{\omega^2}{c^2}\alpha(\omega)\sum_{i\neq j}G(\textbf{r}_i,\textbf{r}_j;\omega)E_j
\end{equation}
where $\textbf{r}_i$ is the position of the scatterer $i$ and $G(\textbf{r},\textbf{r}_p;\omega)$ is the Green function\,(\ref{Green}) \cite{Caze}. Equation\,(\ref{GreenEscat}) allows one to evaluate the scattered field at $\textbf{r}_p$ when the system is illuminated by a source dipole $p=1/\mu_0\omega^2$ located at $\textbf{r}_p$ and oriented along the $\hat{z}$-direction. Once the exciting field at $\textbf{r}_p$ is known, the Purcell spectrum $P(\textbf{r}_p;\omega)$ in the dipole approximation (DA) can be evaluated through the formula $4(1/4 + \Im[E(r_p,\omega)])/\omega^2\mu_0p$, where $\Im[E(r_p,\omega)]$ is the imaginary part of the scattered field at the position of the excitation~\cite{TPSEjosab,AsatryanPRE}.

\begin{figure}[b!h!]
    \includegraphics[width = \linewidth]{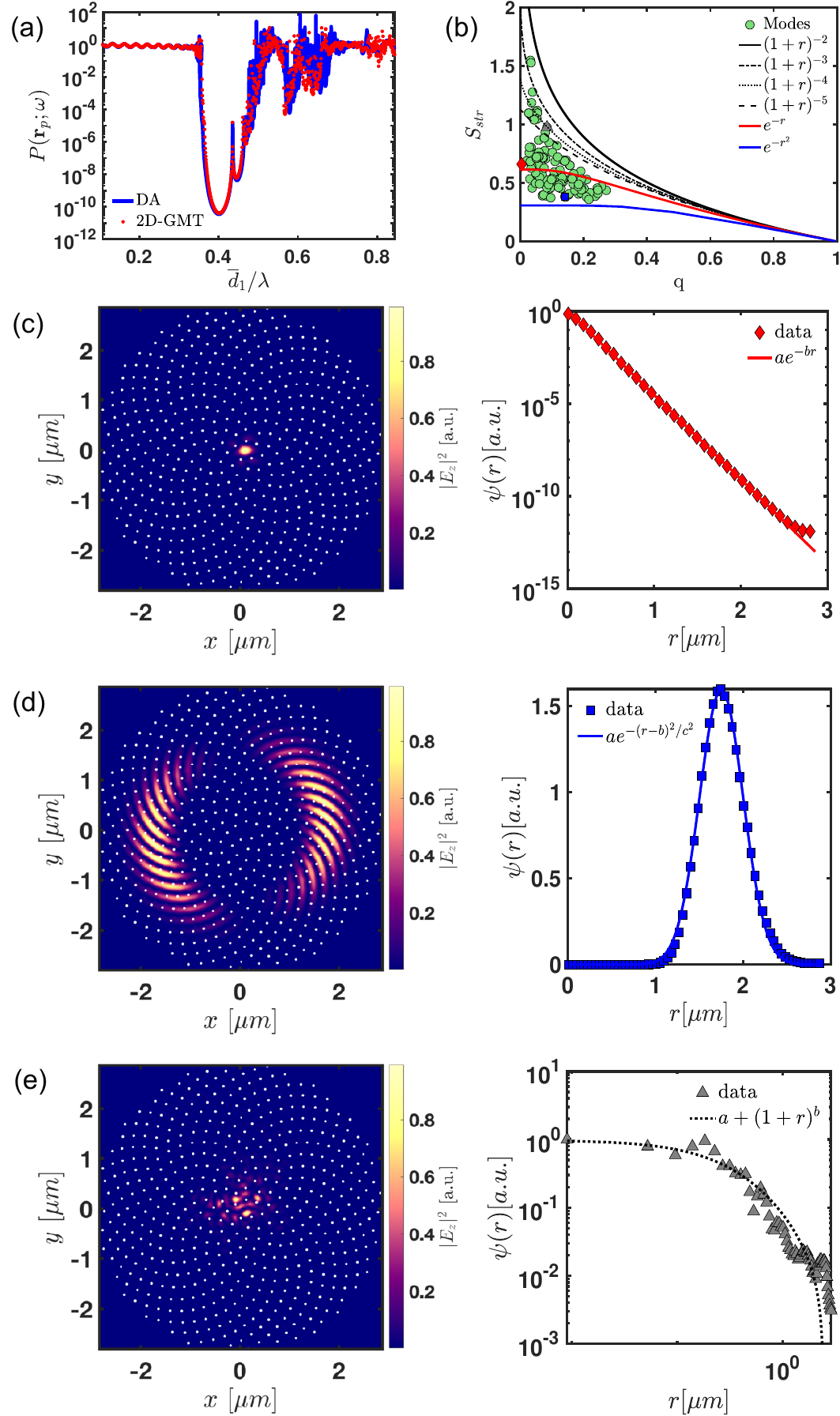}
    \caption{Panel (a) shows the Purcell spectrum of a Golden angle vogel spiral as a function of the averaged interparticle separation $\overline{d_1}$ normalized to the incident wavelength $\lambda$ calculated with the 2D-GMT [red dots] and by solving Eq.\,(\ref{GreenEscat}) [blue line]. Panel (b) displays the structural entropy $S_{str}$ as a function of the filling factor $q$ for all the optical modes corresponding to the different Purcell enhancement peaks of panel (a). Three representative mode profiles (exponential, Gaussian, and power-law) are reported in the left panels of (c), (d), and (e), respectively. The right panels show instead how these modes decay in space by performing an azimuthal average operation. Here, $r$ is equal to $\sqrt{x^2 + y^2}$.}
    \label{2DGMT}
\end{figure}

In Fig.~\ref{2DGMT}(a) we show the Purcell spectra of a golden angle Vogel spiral obtained by using both the 2D-GMT method [red dots] and by solving Eq.\,(\ref{GreenEscat}) [blue line]. In particular, our findings show that  $\omega_0=3.67\times10^{15}\,\text{s}^{-1}$ and $\Gamma_0=9\times10^{15}\,\text{s}^{-1}$ correspond to arrays of dielectric nanocylinder ($\varepsilon=10.5$) of almost 30\,nm radius with an averaged interparticle separation $\overline{d}_1$ equals to 200\,nm. This choice of material, particle size, and averaged interparticle separation guarantees the validity of the electric DA within the 2D-GMT framework. 
By employing the method of Sec.~\ref{Mode2DGMT}, we have evaluated the resonant modes corresponding to the different Purcell enhancement peaks of Fig.~\ref{2DGMT}(a). Specifically, we have evaluated the TM-polarized (i.e., the $\hat{z}$ component of the electric field) resonant modes by generating two-dimensional maps of $\det[\textbf{T}(\tilde{k})]$ with a very large spectral resolution around the narrow Purcell factor peaks, shown in Fig.~\ref{2DGMT}(a). We have used a resolution of $\Delta\Re(\tilde{k})$ equal to $1.3\times10^{-3}\mu m^{-1}$ and a spacing in log-space of $\Delta[\log_{10}\Im(\tilde{k})]$ equal to 0.06. The resulting electromagnetic quasimodes agree very well with what have been previously reported based on both 2D finite element method simulations \cite{TrevinoMF,liew2011localized} and Green's matrix method \cite{sgrignuoli2019localization}. In order to evaluate the structural entropy and therefore the localization maps in the present case, we have used a suitable generalization of Eqs.\,(\ref{qn}) and (\ref{str}) to a continuous distribution \cite{VargaRenyi}. Figure \ref{2DGMT}(b) displays the result of this analysis performed on a total of 65 different resonant modes. Differently from the results of Fig.~\ref{fig:locmap_highOD}, which considered all the scattering resonances of the system, Fig.~\ref{2DGMT}(b) shows the structural entropy $S_{str}$ as a function of $q$ for the resonant modes that can be excited when the material and geometrical parameters of the cylinders are explicitly taken into account. Interestingly, Fig.~\ref{2DGMT}(b) displays the same behavior of Fig.~\ref{fig:locmap_highOD}, demonstrating that optical modes with different types of spatial decay can actually be excited in Vogel spirals. The qualitative agreement between the localization maps obtained via the Green's matrix method and the more general 2D-GMT shows that the former captures the essential spectral properties associated with the aperiodic geometry of the structures, regardless of their specific material parameters.  Additionally, the latter analysis confirms that we can subdivide the modes of a golden angle Vogel spiral into three broad categories, depending on the nature of their spatial decay. In particular, we have identified optical modes with exponential, Gaussian, and power-law decay. Representative examples of each mode category are reported in panels (c-e) of Fig.~\ref{2DGMT}, respectively, along with the corresponding spatial decay curves. Therefore our results demonstrate that the localization map analysis can correctly predict the type of spatial decay of the optical modes of Vogel spiral arrays. 

Moreover, Fig.~\ref{2DGMT} shows the usefulness and applicability of the localization maps in the investigation of the average spatial decay of quasimodes supported by Vogel spirals. Indeed, without using localization maps, if one wanted to determine the spatial decay of optical quasimodes, it would be necessary to analyze the behaviour of every single quasimode supported by the structure, most likely by fitting the spatial profile of each mode, as it has been done in Fig.~\ref{2DGMT}(c-e), right panel. In typical photonic structures, such as the ones investigated in this paper that supports thousands of electromagnetic modes, this is clearly not feasible and, more importantly, this would not capture the global trend that characterise the spatial decay profile of the quasimodes. As a result, the localization map analysis  not only allows one to investigate the overall spatial decay of quasimodes of Vogel spirals, which has never been studied so far, but also provides a unique theoretical tool without which this goal would be very difficult to achieve.

%%%%%%%%%%%%%%%%%%%%%%%%%%%%%%%%%%%%%%%%%%%%%%%%%%%%%%%%%%
%%%%%%%%%%%%%%%%%%		        Conclusions              		   %%%%%%%%%%%%%%%%%%%%
%%%%%%%%%%%%%%%%%%%%%%%%%%%%%%%%%%%%%%%%%%%%%%%%%%%%%%%%%%
\section{Conclusions}

In conclusion, we have studied the spatial decay of quasimodes of several Vogel spiral arrays of electric dipoles in two-dimensions using the concepts of structural entropy and localization maps. The quasimodes are determined via the Green's matrix method as well as the more general 2D-GMT in the dipole limit,  enabling to understand the connection between the spatial decay and the lifetime of the quasimodes. By means of this analysis we find that three types of spatially decaying quasimodes coexist in Vogel spirals, namely exponential, power-law, and Gaussian, which have been independently demonstrated by calculating the electric field in real space. Our analysis shows that the most localized modes in Vogel spirals, both spatially (smallest participation ratios) and temporarily (largest lifetimes), are on average the ones with Gaussian decay. We also provide a direct demonstration of critical modes in Vogel spirals, which are the modes that exhibit power-law decay. Our findings unveil the rich spectrum of optical resonances supported by Vogel spirals structures, with a broad range of coexisting quasimodes that exhibit different types of spatial decay. 

%%%%%%%%%%%%%%%%%%%%%%%%%%%%%%%%%%%%%%%%%%%%%%%%%%%%%%%%%%
%%%%%%%%%%%%%%%%%%		        Acknowledgements              		   %%%%%%%%%%%%%%%%%%%%
%%%%%%%%%%%%%%%%%%%%%%%%%%%%%%%%%%%%%%%%%%%%%%%%%%%%%%%%%%
\section*{Acknowledgements}
We thank R. Bachelard for useful discussions. M. P. and F.A.P acknowledge CNPq, CAPES, and FAPERJ for financial support. L.D.N. acknowledges the support from the Army Research Laboratory under Cooperative Agreement Number W911NF-12-2-0023.

%%%%%%%%%%%%%%%%%%%%%%%%%%%%%%%%%%%%%%%%%%%%%%%%%%%%%%%%%%
%%%%%%%%%%%%%%%%%%		        Biblio              		   %%%%%%%%%%%%%%%%%%%%
%%%%%%%%%%%%%%%%%%%%%%%%%%%%%%%%%%%%%%%%%%%%%%%%%%%%%%%%%%
% \bibliographystyle{apsrev4-1}
% \bibliography{references} %Produces the bibliography via BibTeX.
\input{bibliography.bbl}
%%
%%%%%%%%%%%%%%%%%%%%%%%%%%%%%%%%%%%%%%%%%%%%%%%%%%%%%%%%%%
%%%%%%%%%%%%%%%%%%		        Figure1              		   %%%%%%%%%%%%%%%%%%%%
%%%%%%%%%%%%%%%%%%%%%%%%%%%%%%%%%%%%%%%%%%%%%%%%%%%%%%%%%%
%%%%%%%%%%%%%%%%%%%%%%%%%%%%%%%%%%%%%%%%%%%%%%%%%%%%%%%%%%
%%%%%%%%%%%%%%%%%%		        Figure2           		   %%%%%%%%%%%%%%%%%%%%
%%%%%%%%%%%%%%%%%%%%%%%%%%%%%%%%%%%%%%%%%%%%%%%%%%%%%%%%%%
%%%%%%%%%%%%%%%%%%%%%%%%%%%%%%%%%%%%%%%%%%%%%%%%%%%%%%%%%%

%%%%%%%%%%%%%%%%%%		        Figure3  %%%%%%%%%%%%%%%%%%%%
%%%%%%%%%%%%%%%%%%%%%%%%%%%%%%%%%%%%%%%%%%%%%%%%%%%%%%%%%%
%%%%%%%%%%%%%%%%%%%%%%%%%%%%%%%%%%%%%%%%%%%%%%%%%%%%%%%%%%
\end{document}

%% file: bibliography.bbl

%% file: manuscript.bbl
\begin{thebibliography}{63}%
\makeatletter
\providecommand \@ifxundefined [1]{%
 \@ifx{#1\undefined}
}%
\providecommand \@ifnum [1]{%
 \ifnum #1\expandafter \@firstoftwo
 \else \expandafter \@secondoftwo
 \fi
}%
\providecommand \@ifx [1]{%
 \ifx #1\expandafter \@firstoftwo
 \else \expandafter \@secondoftwo
 \fi
}%
\providecommand \natexlab [1]{#1}%
\providecommand \emph  [1]{``#1''}%
\providecommand \bibnamefont  [1]{#1}%
\providecommand \bibfnamefont [1]{#1}%
\providecommand \citenamefont [1]{#1}%
\providecommand \href@noop [0]{\@secondoftwo}%
\providecommand \href [0]{\begingroup \@sanitize@url \@href}%
\providecommand \@href[1]{\@@startlink{#1}\@@href}%
\providecommand \@@href[1]{\endgroup#1\@@endlink}%
\providecommand \@sanitize@url [0]{\catcode `\\12\catcode `\$12\catcode
  `\&12\catcode `\#12\catcode `\^12\catcode `\_12\catcode `\%12\relax}%
\providecommand \@@startlink[1]{}%
\providecommand \@@endlink[0]{}%
\providecommand \url  [0]{\begingroup\@sanitize@url \@url }%
\providecommand \@url [1]{\endgroup\@href {#1}{\urlprefix }}%
\providecommand \urlprefix  [0]{URL }%
\providecommand \Eprint [0]{\href }%
\providecommand \doibase [0]{http://dx.doi.org/}%
\providecommand \selectlanguage [0]{\@gobble}%
\providecommand \bibinfo  [0]{\@secondoftwo}%
\providecommand \bibfield  [0]{\@secondoftwo}%
\providecommand \translation [1]{[#1]}%
\providecommand \BibitemOpen [0]{}%
\providecommand \bibitemStop [0]{}%
\providecommand \bibitemNoStop [0]{.\EOS\space}%
\providecommand \EOS [0]{\spacefactor3000\relax}%
\providecommand \BibitemShut  [1]{\csname bibitem#1\endcsname}%
\let\auto@bib@innerbib\@empty
%</preamble>
\bibitem [{\citenamefont {Zheludev}\ and\ \citenamefont
  {Kivshar}(2012)}]{zheludev2012metamaterials}%
  \BibitemOpen
  \bibfield  {author} {\bibinfo {author} {\bibfnamefont {N.~I.}\ \bibnamefont
  {Zheludev}}\ and\ \bibinfo {author} {\bibfnamefont {Y.~S.}\ \bibnamefont
  {Kivshar}},\ }\bibfield  {title} {\emph {\bibinfo {title} {From metamaterials
  to metadevices},}\ }\href@noop {} {\bibfield  {journal} {\bibinfo  {journal}
  {Nature materials}\ }\textbf {\bibinfo {volume} {11}},\ \bibinfo {pages}
  {917} (\bibinfo {year} {2012})}\BibitemShut {NoStop}%
\bibitem [{\citenamefont {Segev}\ \emph {et~al.}(2013)\citenamefont {Segev},
  \citenamefont {Silberberg},\ and\ \citenamefont
  {Christodoulides}}]{segev2013anderson}%
  \BibitemOpen
  \bibfield  {author} {\bibinfo {author} {\bibfnamefont {M.}~\bibnamefont
  {Segev}}, \bibinfo {author} {\bibfnamefont {Y.}~\bibnamefont {Silberberg}}, \
  and\ \bibinfo {author} {\bibfnamefont {D.~N.}\ \bibnamefont
  {Christodoulides}},\ }\bibfield  {title} {\emph {\bibinfo {title} {Anderson
  localization of light},}\ }\href@noop {} {\bibfield  {journal} {\bibinfo
  {journal} {Nature Photonics}\ }\textbf {\bibinfo {volume} {7}},\ \bibinfo
  {pages} {197} (\bibinfo {year} {2013})}\BibitemShut {NoStop}%
\bibitem [{\citenamefont {Lagendijk}\ \emph {et~al.}(2009)\citenamefont
  {Lagendijk}, \citenamefont {Van~Tiggelen},\ and\ \citenamefont
  {Wiersma}}]{lagendijk2009fifty}%
  \BibitemOpen
  \bibfield  {author} {\bibinfo {author} {\bibfnamefont {A.}~\bibnamefont
  {Lagendijk}}, \bibinfo {author} {\bibfnamefont {B.}~\bibnamefont
  {Van~Tiggelen}}, \ and\ \bibinfo {author} {\bibfnamefont {D.~S.}\
  \bibnamefont {Wiersma}},\ }\bibfield  {title} {\emph {\bibinfo {title} {Fifty
  years of Anderson localization},}\ }\href@noop {} {\bibfield  {journal}
  {\bibinfo  {journal} {Physics Today}\ }\textbf {\bibinfo {volume} {62}},\
  \bibinfo {pages} {24} (\bibinfo {year} {2009})}\BibitemShut {NoStop}%
\bibitem [{\citenamefont {Wiersma}(2008)}]{wiersma2008physics}%
  \BibitemOpen
  \bibfield  {author} {\bibinfo {author} {\bibfnamefont {D.~S.}\ \bibnamefont
  {Wiersma}},\ }\bibfield  {title} {\emph {\bibinfo {title} {The physics and
  applications of random lasers},}\ }\href@noop {} {\bibfield  {journal}
  {\bibinfo  {journal} {Nature physics}\ }\textbf {\bibinfo {volume} {4}},\
  \bibinfo {pages} {359} (\bibinfo {year} {2008})}\BibitemShut {NoStop}%
\bibitem [{\citenamefont {Gopinath}\ \emph {et~al.}(2009)\citenamefont
  {Gopinath}, \citenamefont {Boriskina}, \citenamefont {Reinhard},\ and\
  \citenamefont {Dal~Negro}}]{Gopinath}%
  \BibitemOpen
  \bibfield  {author} {\bibinfo {author} {\bibfnamefont {A.}~\bibnamefont
  {Gopinath}}, \bibinfo {author} {\bibfnamefont {S.~V.}\ \bibnamefont
  {Boriskina}}, \bibinfo {author} {\bibfnamefont {B.~M.}\ \bibnamefont
  {Reinhard}}, \ and\ \bibinfo {author} {\bibfnamefont {L.}~\bibnamefont
  {Dal~Negro}},\ }\bibfield  {title} {\emph {\bibinfo {title} {Deterministic
  aperiodic arrays of metal nanoparticles for surface-enhanced Raman scattering
  (SERS)},}\ }\href@noop {} {\bibfield  {journal} {\bibinfo  {journal} {Optics
  Express}\ }\textbf {\bibinfo {volume} {17}},\ \bibinfo {pages} {3741}
  (\bibinfo {year} {2009})}\BibitemShut {NoStop}%
\bibitem [{\citenamefont {Lee}\ \emph {et~al.}(2010)\citenamefont {Lee},
  \citenamefont {Amsden}, \citenamefont {Boriskina}, \citenamefont {Gopinath},
  \citenamefont {Mitropolous}, \citenamefont {Kaplan}, \citenamefont
  {Omenetto},\ and\ \citenamefont {Dal~Negro}}]{Lee}%
  \BibitemOpen
  \bibfield  {author} {\bibinfo {author} {\bibfnamefont {S.~Y.}\ \bibnamefont
  {Lee}}, \bibinfo {author} {\bibfnamefont {J.~J.}\ \bibnamefont {Amsden}},
  \bibinfo {author} {\bibfnamefont {S.~V.}\ \bibnamefont {Boriskina}}, \bibinfo
  {author} {\bibfnamefont {A.}~\bibnamefont {Gopinath}}, \bibinfo {author}
  {\bibfnamefont {A.}~\bibnamefont {Mitropolous}}, \bibinfo {author}
  {\bibfnamefont {D.~L.}\ \bibnamefont {Kaplan}}, \bibinfo {author}
  {\bibfnamefont {F.~G.}\ \bibnamefont {Omenetto}}, \ and\ \bibinfo {author}
  {\bibfnamefont {L.}~\bibnamefont {Dal~Negro}},\ }\bibfield  {title} {\emph
  {\bibinfo {title} {Spatial and spectral detection of protein monolayers with
  deterministic aperiodic arrays of metal nanoparticles},}\ }\href@noop {}
  {\bibfield  {journal} {\bibinfo  {journal} {Proceedings of the National
  Academy of Sciences}\ }\textbf {\bibinfo {volume} {107}},\ \bibinfo {pages}
  {12086} (\bibinfo {year} {2010})}\BibitemShut {NoStop}%
\bibitem [{\citenamefont {Noh}\ \emph {et~al.}(2011)\citenamefont {Noh},
  \citenamefont {Yang}, \citenamefont {Boriskina}, \citenamefont {Rooks},
  \citenamefont {Solomon}, \citenamefont {Dal~Negro},\ and\ \citenamefont
  {Cao}}]{noh2011lasing}%
  \BibitemOpen
  \bibfield  {author} {\bibinfo {author} {\bibfnamefont {H.}~\bibnamefont
  {Noh}}, \bibinfo {author} {\bibfnamefont {J.-K.}\ \bibnamefont {Yang}},
  \bibinfo {author} {\bibfnamefont {S.~V.}\ \bibnamefont {Boriskina}}, \bibinfo
  {author} {\bibfnamefont {M.~J.}\ \bibnamefont {Rooks}}, \bibinfo {author}
  {\bibfnamefont {G.~S.}\ \bibnamefont {Solomon}}, \bibinfo {author}
  {\bibfnamefont {L.}~\bibnamefont {Dal~Negro}}, \ and\ \bibinfo {author}
  {\bibfnamefont {H.}~\bibnamefont {Cao}},\ }\bibfield  {title} {\emph
  {\bibinfo {title} {Lasing in Thue--Morse structures with optimized
  aperiodicity},}\ }\href@noop {} {\bibfield  {journal} {\bibinfo  {journal}
  {Applied Physics Letters}\ }\textbf {\bibinfo {volume} {98}},\ \bibinfo
  {pages} {201109} (\bibinfo {year} {2011})}\BibitemShut {NoStop}%
\bibitem [{\citenamefont {Dal~Negro}\ and\ \citenamefont
  {Boriskina}(2012)}]{DalNegroReview}%
  \BibitemOpen
  \bibfield  {author} {\bibinfo {author} {\bibfnamefont {L.}~\bibnamefont
  {Dal~Negro}}\ and\ \bibinfo {author} {\bibfnamefont {S.~V.}\ \bibnamefont
  {Boriskina}},\ }\bibfield  {title} {\emph {\bibinfo {title} {Deterministic
  aperiodic nanostructures for photonics and plasmonics applications},}\
  }\href@noop {} {\bibfield  {journal} {\bibinfo  {journal} {Laser \& Photonics
  Reviews}\ }\textbf {\bibinfo {volume} {6}},\ \bibinfo {pages} {178} (\bibinfo
  {year} {2012})}\BibitemShut {NoStop}%
\bibitem [{\citenamefont {Dal~Negro}(2013)}]{DalNegroBook}%
  \BibitemOpen
  \bibfield  {author} {\bibinfo {author} {\bibfnamefont {L.}~\bibnamefont
  {Dal~Negro}},\ }\href@noop {} {\emph {\bibinfo {title} {Optics of aperiodic
  structures: fundamentals and device applications}}}\ (\bibinfo  {publisher}
  {CRC Press},\ \bibinfo {year} {2013})\BibitemShut {NoStop}%
\bibitem [{\citenamefont {Maci{\'a}}(2008)}]{MaciaBook}%
  \BibitemOpen
  \bibfield  {author} {\bibinfo {author} {\bibfnamefont {E.}~\bibnamefont
  {Maci{\'a}}},\ }\href@noop {} {\emph {\bibinfo {title} {Aperiodic structures
  in condensed matter: fundamentals and applications}}}\ (\bibinfo  {publisher}
  {CRC Press},\ \bibinfo {year} {2008})\BibitemShut {NoStop}%
\bibitem [{\citenamefont {Maci{\'a}-Barber}(2020)}]{MaciaBook2}%
  \BibitemOpen
  \bibfield  {author} {\bibinfo {author} {\bibfnamefont {E.}~\bibnamefont
  {Maci{\'a}-Barber}},\ }\href@noop {} {\emph {\bibinfo {title} {Quasicrystals:
  Fundamentals and Applications}}}\ (\bibinfo  {publisher} {CRC Press},\
  \bibinfo {year} {2020})\BibitemShut {NoStop}%
\bibitem [{\citenamefont {Dal~Negro}\ \emph {et~al.}(2016)\citenamefont
  {Dal~Negro}, \citenamefont {Wang},\ and\ \citenamefont
  {Pinheiro}}]{DalNegroCrystals}%
  \BibitemOpen
  \bibfield  {author} {\bibinfo {author} {\bibfnamefont {L.}~\bibnamefont
  {Dal~Negro}}, \bibinfo {author} {\bibfnamefont {R.}~\bibnamefont {Wang}}, \
  and\ \bibinfo {author} {\bibfnamefont {F.~A.}\ \bibnamefont {Pinheiro}},\
  }\bibfield  {title} {\emph {\bibinfo {title} {Structural and spectral
  properties of deterministic aperiodic optical structures},}\ }\href@noop {}
  {\bibfield  {journal} {\bibinfo  {journal} {Crystals}\ }\textbf {\bibinfo
  {volume} {6}},\ \bibinfo {pages} {161} (\bibinfo {year} {2016})}\BibitemShut
  {NoStop}%
\bibitem [{\citenamefont {Sgrignuoli}\ \emph {et~al.}(2020)\citenamefont
  {Sgrignuoli}, \citenamefont {Gorsky}, \citenamefont {Britton}, \citenamefont
  {Zhang}, \citenamefont {Riboli},\ and\ \citenamefont
  {Dal~Negro}}]{sgrignuoli2020multifractality}%
  \BibitemOpen
  \bibfield  {author} {\bibinfo {author} {\bibfnamefont {F.}~\bibnamefont
  {Sgrignuoli}}, \bibinfo {author} {\bibfnamefont {S.}~\bibnamefont {Gorsky}},
  \bibinfo {author} {\bibfnamefont {W.~A.}\ \bibnamefont {Britton}}, \bibinfo
  {author} {\bibfnamefont {R.}~\bibnamefont {Zhang}}, \bibinfo {author}
  {\bibfnamefont {F.}~\bibnamefont {Riboli}}, \ and\ \bibinfo {author}
  {\bibfnamefont {L.}~\bibnamefont {Dal~Negro}},\ }\bibfield  {title} {\emph
  {\bibinfo {title} {Multifractality of light in photonic arrays based on
  algebraic number theory},}\ }\href@noop {} {\bibfield  {journal} {\bibinfo
  {journal} {Communications Physics}\ }\textbf {\bibinfo {volume} {3}},\
  \bibinfo {pages} {106} (\bibinfo {year} {2020})}\BibitemShut {NoStop}%
\bibitem [{\citenamefont {Sgrignuoli}\ and\ \citenamefont
  {Dal~Negro}(2020)}]{sgrignuoli2020subdiffusive}%
  \BibitemOpen
  \bibfield  {author} {\bibinfo {author} {\bibfnamefont {F.}~\bibnamefont
  {Sgrignuoli}}\ and\ \bibinfo {author} {\bibfnamefont {L.}~\bibnamefont
  {Dal~Negro}},\ }\bibfield  {title} {\emph {\bibinfo {title} {Subdiffusive
  light transport in three-dimensional subrandom arrays},}\ }\href@noop {}
  {\bibfield  {journal} {\bibinfo  {journal} {Physical Review B}\ }\textbf
  {\bibinfo {volume} {101}},\ \bibinfo {pages} {214204} (\bibinfo {year}
  {2020})}\BibitemShut {NoStop}%
\bibitem [{\citenamefont {Negro}\ \emph {et~al.}(2021)\citenamefont {Negro},
  \citenamefont {Chen}, \citenamefont {Gorsky},\ and\ \citenamefont
  {Sgrignuoli}}]{TPSEjosab}%
  \BibitemOpen
  \bibfield  {author} {\bibinfo {author} {\bibfnamefont {L.~D.}\ \bibnamefont
  {Negro}}, \bibinfo {author} {\bibfnamefont {Y.}~\bibnamefont {Chen}},
  \bibinfo {author} {\bibfnamefont {S.}~\bibnamefont {Gorsky}}, \ and\ \bibinfo
  {author} {\bibfnamefont {F.}~\bibnamefont {Sgrignuoli}},\ }\bibfield  {title}
  {\emph {\bibinfo {title} {Aperiodic bandgap structures for enhanced quantum
  two-photon sources},}\ }\href@noop {} {\bibfield  {journal} {\bibinfo
  {journal} {Journal of the Optical Society of America B}\ }\textbf {\bibinfo
  {volume} {38}},\ \bibinfo {pages} {C94} (\bibinfo {year} {2021})}\BibitemShut
  {NoStop}%
\bibitem [{\citenamefont {Sgrignuoli}\ \emph {et~al.}(2019)\citenamefont
  {Sgrignuoli}, \citenamefont {Wang}, \citenamefont {Pinheiro},\ and\
  \citenamefont {Dal~Negro}}]{sgrignuoli2019localization}%
  \BibitemOpen
  \bibfield  {author} {\bibinfo {author} {\bibfnamefont {F.}~\bibnamefont
  {Sgrignuoli}}, \bibinfo {author} {\bibfnamefont {R.}~\bibnamefont {Wang}},
  \bibinfo {author} {\bibfnamefont {F.}~\bibnamefont {Pinheiro}}, \ and\
  \bibinfo {author} {\bibfnamefont {L.}~\bibnamefont {Dal~Negro}},\ }\bibfield
  {title} {\emph {\bibinfo {title} {Localization of scattering resonances in
  aperiodic Vogel spirals},}\ }\href@noop {} {\bibfield  {journal} {\bibinfo
  {journal} {Physical Review B}\ }\textbf {\bibinfo {volume} {99}},\ \bibinfo
  {pages} {104202} (\bibinfo {year} {2019})}\BibitemShut {NoStop}%
\bibitem [{\citenamefont {Vardeny}\ \emph {et~al.}(2013)\citenamefont
  {Vardeny}, \citenamefont {Nahata},\ and\ \citenamefont
  {Agrawal}}]{vardeny2013optics}%
  \BibitemOpen
  \bibfield  {author} {\bibinfo {author} {\bibfnamefont {Z.~V.}\ \bibnamefont
  {Vardeny}}, \bibinfo {author} {\bibfnamefont {A.}~\bibnamefont {Nahata}}, \
  and\ \bibinfo {author} {\bibfnamefont {A.}~\bibnamefont {Agrawal}},\
  }\bibfield  {title} {\emph {\bibinfo {title} {Optics of photonic
  quasicrystals},}\ }\href@noop {} {\bibfield  {journal} {\bibinfo  {journal}
  {Nature photonics}\ }\textbf {\bibinfo {volume} {7}},\ \bibinfo {pages} {177}
  (\bibinfo {year} {2013})}\BibitemShut {NoStop}%
\bibitem [{\citenamefont {Razi}\ \emph {et~al.}(2019)\citenamefont {Razi},
  \citenamefont {Wang}, \citenamefont {He}, \citenamefont {Kirby},\ and\
  \citenamefont {Dal~Negro}}]{razi2019optimization}%
  \BibitemOpen
  \bibfield  {author} {\bibinfo {author} {\bibfnamefont {M.}~\bibnamefont
  {Razi}}, \bibinfo {author} {\bibfnamefont {R.}~\bibnamefont {Wang}}, \bibinfo
  {author} {\bibfnamefont {Y.}~\bibnamefont {He}}, \bibinfo {author}
  {\bibfnamefont {R.~M.}\ \bibnamefont {Kirby}}, \ and\ \bibinfo {author}
  {\bibfnamefont {L.}~\bibnamefont {Dal~Negro}},\ }\bibfield  {title} {\emph
  {\bibinfo {title} {{Optimization of large-scale Vogel spiral arrays of
  plasmonic nanoparticles}},}\ }\href@noop {} {\bibfield  {journal} {\bibinfo
  {journal} {Plasmonics}\ }\textbf {\bibinfo {volume} {14}},\ \bibinfo {pages}
  {253} (\bibinfo {year} {2019})}\BibitemShut {NoStop}%
\bibitem [{\citenamefont {Trevino}\ \emph {et~al.}(2011)\citenamefont
  {Trevino}, \citenamefont {Cao},\ and\ \citenamefont
  {Dal~Negro}}]{trevino2011circularly}%
  \BibitemOpen
  \bibfield  {author} {\bibinfo {author} {\bibfnamefont {J.}~\bibnamefont
  {Trevino}}, \bibinfo {author} {\bibfnamefont {H.}~\bibnamefont {Cao}}, \ and\
  \bibinfo {author} {\bibfnamefont {L.}~\bibnamefont {Dal~Negro}},\ }\bibfield
  {title} {\emph {\bibinfo {title} {Circularly symmetric light scattering from
  nanoplasmonic spirals},}\ }\href@noop {} {\bibfield  {journal} {\bibinfo
  {journal} {Nano letters}\ }\textbf {\bibinfo {volume} {11}},\ \bibinfo
  {pages} {2008} (\bibinfo {year} {2011})}\BibitemShut {NoStop}%
\bibitem [{\citenamefont {Huang}\ \emph {et~al.}(2007)\citenamefont {Huang},
  \citenamefont {Chen}, \citenamefont {de~Abajo},\ and\ \citenamefont
  {Zheludev}}]{huang2007optical}%
  \BibitemOpen
  \bibfield  {author} {\bibinfo {author} {\bibfnamefont {F.~M.}\ \bibnamefont
  {Huang}}, \bibinfo {author} {\bibfnamefont {Y.}~\bibnamefont {Chen}},
  \bibinfo {author} {\bibfnamefont {F.~J.~G.}\ \bibnamefont {de~Abajo}}, \ and\
  \bibinfo {author} {\bibfnamefont {N.~I.}\ \bibnamefont {Zheludev}},\
  }\bibfield  {title} {\emph {\bibinfo {title} {Optical super-resolution
  through super-oscillations},}\ }\href@noop {} {\bibfield  {journal} {\bibinfo
   {journal} {Journal of Optics A: Pure and Applied Optics}\ }\textbf {\bibinfo
  {volume} {9}},\ \bibinfo {pages} {S285} (\bibinfo {year} {2007})}\BibitemShut
  {NoStop}%
\bibitem [{\citenamefont {Dal~Negro}\ \emph {et~al.}(2012)\citenamefont
  {Dal~Negro}, \citenamefont {Lawrence},\ and\ \citenamefont
  {Trevino}}]{dal2012analytical}%
  \BibitemOpen
  \bibfield  {author} {\bibinfo {author} {\bibfnamefont {L.}~\bibnamefont
  {Dal~Negro}}, \bibinfo {author} {\bibfnamefont {N.}~\bibnamefont {Lawrence}},
  \ and\ \bibinfo {author} {\bibfnamefont {J.}~\bibnamefont {Trevino}},\
  }\bibfield  {title} {\emph {\bibinfo {title} {Analytical light scattering and
  orbital angular momentum spectra of arbitrary Vogel spirals},}\ }\href@noop
  {} {\bibfield  {journal} {\bibinfo  {journal} {Optics express}\ }\textbf
  {\bibinfo {volume} {20}},\ \bibinfo {pages} {18209} (\bibinfo {year}
  {2012})}\BibitemShut {NoStop}%
\bibitem [{\citenamefont {Lawrence}\ \emph
  {et~al.}(2012{\natexlab{a}})\citenamefont {Lawrence}, \citenamefont
  {Trevino},\ and\ \citenamefont {Dal~Negro}}]{lawrence2012control}%
  \BibitemOpen
  \bibfield  {author} {\bibinfo {author} {\bibfnamefont {N.}~\bibnamefont
  {Lawrence}}, \bibinfo {author} {\bibfnamefont {J.}~\bibnamefont {Trevino}}, \
  and\ \bibinfo {author} {\bibfnamefont {L.}~\bibnamefont {Dal~Negro}},\
  }\bibfield  {title} {\emph {\bibinfo {title} {Control of optical orbital
  angular momentum by Vogel spiral arrays of metallic nanoparticles},}\
  }\href@noop {} {\bibfield  {journal} {\bibinfo  {journal} {Optics letters}\
  }\textbf {\bibinfo {volume} {37}},\ \bibinfo {pages} {5076} (\bibinfo {year}
  {2012}{\natexlab{a}})}\BibitemShut {NoStop}%
\bibitem [{\citenamefont {Liew}\ \emph {et~al.}(2011)\citenamefont {Liew},
  \citenamefont {Noh}, \citenamefont {Trevino}, \citenamefont {Dal~Negro},\
  and\ \citenamefont {Cao}}]{liew2011localized}%
  \BibitemOpen
  \bibfield  {author} {\bibinfo {author} {\bibfnamefont {S.~F.}\ \bibnamefont
  {Liew}}, \bibinfo {author} {\bibfnamefont {H.}~\bibnamefont {Noh}}, \bibinfo
  {author} {\bibfnamefont {J.}~\bibnamefont {Trevino}}, \bibinfo {author}
  {\bibfnamefont {L.}~\bibnamefont {Dal~Negro}}, \ and\ \bibinfo {author}
  {\bibfnamefont {H.}~\bibnamefont {Cao}},\ }\bibfield  {title} {\emph
  {\bibinfo {title} {Localized photonic band edge modes and orbital angular
  momenta of light in a golden-angle spiral},}\ }\href@noop {} {\bibfield
  {journal} {\bibinfo  {journal} {Optics express}\ }\textbf {\bibinfo {volume}
  {19}},\ \bibinfo {pages} {23631} (\bibinfo {year} {2011})}\BibitemShut
  {NoStop}%
\bibitem [{\citenamefont {Christofi}\ \emph {et~al.}(2016)\citenamefont
  {Christofi}, \citenamefont {Pinheiro},\ and\ \citenamefont
  {Dal~Negro}}]{christofi2016probing}%
  \BibitemOpen
  \bibfield  {author} {\bibinfo {author} {\bibfnamefont {A.}~\bibnamefont
  {Christofi}}, \bibinfo {author} {\bibfnamefont {F.~A.}\ \bibnamefont
  {Pinheiro}}, \ and\ \bibinfo {author} {\bibfnamefont {L.}~\bibnamefont
  {Dal~Negro}},\ }\bibfield  {title} {\emph {\bibinfo {title} {Probing
  scattering resonances of Vogel's spirals with the Green's matrix spectral
  method},}\ }\href@noop {} {\bibfield  {journal} {\bibinfo  {journal} {Optics
  letters}\ }\textbf {\bibinfo {volume} {41}},\ \bibinfo {pages} {1933}
  (\bibinfo {year} {2016})}\BibitemShut {NoStop}%
\bibitem [{\citenamefont {Pollard}\ and\ \citenamefont
  {Parker}(2009)}]{Pollard}%
  \BibitemOpen
  \bibfield  {author} {\bibinfo {author} {\bibfnamefont {M.~E.}\ \bibnamefont
  {Pollard}}\ and\ \bibinfo {author} {\bibfnamefont {G.~J.}\ \bibnamefont
  {Parker}},\ }\bibfield  {title} {\emph {\bibinfo {title} {Low-contrast
  bandgaps of a planar parabolic spiral lattice},}\ }\href@noop {} {\bibfield
  {journal} {\bibinfo  {journal} {Optics letters}\ }\textbf {\bibinfo {volume}
  {34}},\ \bibinfo {pages} {2805} (\bibinfo {year} {2009})}\BibitemShut
  {NoStop}%
\bibitem [{\citenamefont {Naylor}(2002)}]{Naylor}%
  \BibitemOpen
  \bibfield  {author} {\bibinfo {author} {\bibfnamefont {M.}~\bibnamefont
  {Naylor}},\ }\bibfield  {title} {\emph {\bibinfo {title} {Golden, and $\pi$
  Flowers: A Spiral Story},}\ }\href@noop {} {\bibfield  {journal} {\bibinfo
  {journal} {Mathematics Magazine}\ }\textbf {\bibinfo {volume} {75}},\
  \bibinfo {pages} {163} (\bibinfo {year} {2002})}\BibitemShut {NoStop}%
\bibitem [{\citenamefont {Ryu}\ \emph {et~al.}(1992)\citenamefont {Ryu},
  \citenamefont {Oh},\ and\ \citenamefont {Lee}}]{Ryu1992}%
  \BibitemOpen
  \bibfield  {author} {\bibinfo {author} {\bibfnamefont {C.~S.}\ \bibnamefont
  {Ryu}}, \bibinfo {author} {\bibfnamefont {G.~Y.}\ \bibnamefont {Oh}}, \ and\
  \bibinfo {author} {\bibfnamefont {M.~H.}\ \bibnamefont {Lee}},\ }\bibfield
  {title} {\emph {\bibinfo {title} {{Extended and critical wave functions in a
  Thue-Morse chain}},}\ }\href@noop {} {\bibfield  {journal} {\bibinfo
  {journal} {Physical Review B}\ }\textbf {\bibinfo {volume} {46}},\ \bibinfo
  {pages} {5162} (\bibinfo {year} {1992})}\BibitemShut {NoStop}%
\bibitem [{\citenamefont {Desideri}\ \emph {et~al.}(1989)\citenamefont
  {Desideri}, \citenamefont {Macon},\ and\ \citenamefont
  {Sornette}}]{Desideri1989}%
  \BibitemOpen
  \bibfield  {author} {\bibinfo {author} {\bibfnamefont {J.~P.}\ \bibnamefont
  {Desideri}}, \bibinfo {author} {\bibfnamefont {L.}~\bibnamefont {Macon}}, \
  and\ \bibinfo {author} {\bibfnamefont {D.}~\bibnamefont {Sornette}},\
  }\bibfield  {title} {\emph {\bibinfo {title} {{Observation of critical modes
  in quasiperiodic systems}},}\ }\href@noop {} {\bibfield  {journal} {\bibinfo
  {journal} {Physical Review Letters}\ }\textbf {\bibinfo {volume} {63}},\
  \bibinfo {pages} {390} (\bibinfo {year} {1989})}\BibitemShut {NoStop}%
\bibitem [{\citenamefont {Maci{\'{a}}}(1999)}]{Macia1999}%
  \BibitemOpen
  \bibfield  {author} {\bibinfo {author} {\bibfnamefont {E.}~\bibnamefont
  {Maci{\'{a}}}},\ }\bibfield  {title} {\emph {\bibinfo {title} {{Physical
  nature of critical modes in Fibonacci quasicrystals}},}\ }\href@noop {}
  {\bibfield  {journal} {\bibinfo  {journal} {Physical Review B}\ }\textbf
  {\bibinfo {volume} {60}},\ \bibinfo {pages} {10032} (\bibinfo {year}
  {1999})}\BibitemShut {NoStop}%
\bibitem [{\citenamefont {Mahler}\ \emph {et~al.}(2010)\citenamefont {Mahler},
  \citenamefont {Tredicucci}, \citenamefont {Beltram}, \citenamefont {Walther},
  \citenamefont {Faist}, \citenamefont {Beere}, \citenamefont {Ritchie},\ and\
  \citenamefont {Wiersma}}]{mahler2010quasi}%
  \BibitemOpen
  \bibfield  {author} {\bibinfo {author} {\bibfnamefont {L.}~\bibnamefont
  {Mahler}}, \bibinfo {author} {\bibfnamefont {A.}~\bibnamefont {Tredicucci}},
  \bibinfo {author} {\bibfnamefont {F.}~\bibnamefont {Beltram}}, \bibinfo
  {author} {\bibfnamefont {C.}~\bibnamefont {Walther}}, \bibinfo {author}
  {\bibfnamefont {J.}~\bibnamefont {Faist}}, \bibinfo {author} {\bibfnamefont
  {H.~E.}\ \bibnamefont {Beere}}, \bibinfo {author} {\bibfnamefont {D.~A.}\
  \bibnamefont {Ritchie}}, \ and\ \bibinfo {author} {\bibfnamefont {D.~S.}\
  \bibnamefont {Wiersma}},\ }\bibfield  {title} {\emph {\bibinfo {title}
  {Quasi-periodic distributed feedback laser},}\ }\href@noop {} {\bibfield
  {journal} {\bibinfo  {journal} {Nature Photonics}\ }\textbf {\bibinfo
  {volume} {4}},\ \bibinfo {pages} {165} (\bibinfo {year} {2010})}\BibitemShut
  {NoStop}%
\bibitem [{\citenamefont {Pipek}\ and\ \citenamefont
  {Varga}(1992)}]{Pipek1992}%
  \BibitemOpen
  \bibfield  {author} {\bibinfo {author} {\bibfnamefont {J.}~\bibnamefont
  {Pipek}}\ and\ \bibinfo {author} {\bibfnamefont {I.}~\bibnamefont {Varga}},\
  }\bibfield  {title} {\emph {\bibinfo {title} {Universal classification scheme
  for the spatial-localization properties of one-particle states in finite,
  $d$-dimensional systems},}\ }\href {\doibase 10.1103/PhysRevA.46.3148}
  {\bibfield  {journal} {\bibinfo  {journal} {Physical Review A}\ }\textbf
  {\bibinfo {volume} {46}},\ \bibinfo {pages} {3148} (\bibinfo {year}
  {1992})}\BibitemShut {NoStop}%
\bibitem [{\citenamefont {Santos}\ and\ \citenamefont
  {Rigol}(2010)}]{santos2010localization}%
  \BibitemOpen
  \bibfield  {author} {\bibinfo {author} {\bibfnamefont {L.~F.}\ \bibnamefont
  {Santos}}\ and\ \bibinfo {author} {\bibfnamefont {M.}~\bibnamefont {Rigol}},\
  }\bibfield  {title} {\emph {\bibinfo {title} {Localization and the effects of
  symmetries in the thermalization properties of one-dimensional quantum
  systems},}\ }\href@noop {} {\bibfield  {journal} {\bibinfo  {journal}
  {Physical Review E}\ }\textbf {\bibinfo {volume} {82}},\ \bibinfo {pages}
  {031130} (\bibinfo {year} {2010})}\BibitemShut {NoStop}%
\bibitem [{\citenamefont {Aulbach}\ \emph {et~al.}(2004)\citenamefont
  {Aulbach}, \citenamefont {Wobst}, \citenamefont {Ingold}, \citenamefont
  {H{\"a}nggi},\ and\ \citenamefont {Varga}}]{aulbach2004phase}%
  \BibitemOpen
  \bibfield  {author} {\bibinfo {author} {\bibfnamefont {C.}~\bibnamefont
  {Aulbach}}, \bibinfo {author} {\bibfnamefont {A.}~\bibnamefont {Wobst}},
  \bibinfo {author} {\bibfnamefont {G.-L.}\ \bibnamefont {Ingold}}, \bibinfo
  {author} {\bibfnamefont {P.}~\bibnamefont {H{\"a}nggi}}, \ and\ \bibinfo
  {author} {\bibfnamefont {I.}~\bibnamefont {Varga}},\ }\bibfield  {title}
  {\emph {\bibinfo {title} {Phase-space visualization of a metal--insulator
  transition},}\ }\href@noop {} {\bibfield  {journal} {\bibinfo  {journal} {New
  Journal of Physics}\ }\textbf {\bibinfo {volume} {6}},\ \bibinfo {pages} {70}
  (\bibinfo {year} {2004})}\BibitemShut {NoStop}%
\bibitem [{\citenamefont {Varga}\ \emph {et~al.}(1995)\citenamefont {Varga},
  \citenamefont {Hofstetter}, \citenamefont {Schreiber},\ and\ \citenamefont
  {Pipek}}]{varga1995shape}%
  \BibitemOpen
  \bibfield  {author} {\bibinfo {author} {\bibfnamefont {I.}~\bibnamefont
  {Varga}}, \bibinfo {author} {\bibfnamefont {E.}~\bibnamefont {Hofstetter}},
  \bibinfo {author} {\bibfnamefont {M.}~\bibnamefont {Schreiber}}, \ and\
  \bibinfo {author} {\bibfnamefont {J.}~\bibnamefont {Pipek}},\ }\bibfield
  {title} {\emph {\bibinfo {title} {Shape analysis of the level-spacing
  distribution around the metal-insulator transition in the three-dimensional
  Anderson model},}\ }\href@noop {} {\bibfield  {journal} {\bibinfo  {journal}
  {Physical Review B}\ }\textbf {\bibinfo {volume} {52}},\ \bibinfo {pages}
  {7783} (\bibinfo {year} {1995})}\BibitemShut {NoStop}%
\bibitem [{\citenamefont {Rieth}\ and\ \citenamefont
  {Schreiber}(1998)}]{rieth1998numerical}%
  \BibitemOpen
  \bibfield  {author} {\bibinfo {author} {\bibfnamefont {T.}~\bibnamefont
  {Rieth}}\ and\ \bibinfo {author} {\bibfnamefont {M.}~\bibnamefont
  {Schreiber}},\ }\bibfield  {title} {\emph {\bibinfo {title} {Numerical
  investigation of electronic wave functions in quasiperiodic lattices},}\
  }\href@noop {} {\bibfield  {journal} {\bibinfo  {journal} {Journal of
  Physics: Condensed Matter}\ }\textbf {\bibinfo {volume} {10}},\ \bibinfo
  {pages} {783} (\bibinfo {year} {1998})}\BibitemShut {NoStop}%
\bibitem [{\citenamefont {Rusek}\ \emph {et~al.}(2000)\citenamefont {Rusek},
  \citenamefont {Mostowski},\ and\ \citenamefont
  {Or{\l}owski}}]{rusek2000random}%
  \BibitemOpen
  \bibfield  {author} {\bibinfo {author} {\bibfnamefont {M.}~\bibnamefont
  {Rusek}}, \bibinfo {author} {\bibfnamefont {J.}~\bibnamefont {Mostowski}}, \
  and\ \bibinfo {author} {\bibfnamefont {A.}~\bibnamefont {Or{\l}owski}},\
  }\bibfield  {title} {\emph {\bibinfo {title} {Random Green matrices: From
  proximity resonances to Anderson localization},}\ }\href@noop {} {\bibfield
  {journal} {\bibinfo  {journal} {Physical Review A}\ }\textbf {\bibinfo
  {volume} {61}},\ \bibinfo {pages} {022704} (\bibinfo {year}
  {2000})}\BibitemShut {NoStop}%
\bibitem [{\citenamefont {Pinheiro}\ \emph {et~al.}(2004)\citenamefont
  {Pinheiro}, \citenamefont {Rusek}, \citenamefont {Orlowski},\ and\
  \citenamefont {Van~Tiggelen}}]{pinheiro2004probing}%
  \BibitemOpen
  \bibfield  {author} {\bibinfo {author} {\bibfnamefont {F.}~\bibnamefont
  {Pinheiro}}, \bibinfo {author} {\bibfnamefont {M.}~\bibnamefont {Rusek}},
  \bibinfo {author} {\bibfnamefont {A.}~\bibnamefont {Orlowski}}, \ and\
  \bibinfo {author} {\bibfnamefont {B.}~\bibnamefont {Van~Tiggelen}},\
  }\bibfield  {title} {\emph {\bibinfo {title} {Probing Anderson localization
  of light via decay rate statistics},}\ }\href@noop {} {\bibfield  {journal}
  {\bibinfo  {journal} {Physical Review E}\ }\textbf {\bibinfo {volume} {69}},\
  \bibinfo {pages} {026605} (\bibinfo {year} {2004})}\BibitemShut {NoStop}%
\bibitem [{\citenamefont {Pinheiro}(2008)}]{pinheiro2008statistics}%
  \BibitemOpen
  \bibfield  {author} {\bibinfo {author} {\bibfnamefont {F.}~\bibnamefont
  {Pinheiro}},\ }\bibfield  {title} {\emph {\bibinfo {title} {Statistics of
  quality factors in three-dimensional disordered magneto-optical systems and
  its applications to random lasers},}\ }\href@noop {} {\bibfield  {journal}
  {\bibinfo  {journal} {Physical Review A}\ }\textbf {\bibinfo {volume} {78}},\
  \bibinfo {pages} {023812} (\bibinfo {year} {2008})}\BibitemShut {NoStop}%
\bibitem [{\citenamefont {Goetschy}\ and\ \citenamefont
  {Skipetrov}(2011)}]{goetschy2011non}%
  \BibitemOpen
  \bibfield  {author} {\bibinfo {author} {\bibfnamefont {A.}~\bibnamefont
  {Goetschy}}\ and\ \bibinfo {author} {\bibfnamefont {S.}~\bibnamefont
  {Skipetrov}},\ }\bibfield  {title} {\emph {\bibinfo {title} {Non-Hermitian
  Euclidean random matrix theory},}\ }\href@noop {} {\bibfield  {journal}
  {\bibinfo  {journal} {Physical Review E}\ }\textbf {\bibinfo {volume} {84}},\
  \bibinfo {pages} {011150} (\bibinfo {year} {2011})}\BibitemShut {NoStop}%
\bibitem [{\citenamefont {Skipetrov}\ and\ \citenamefont
  {Goetschy}(2011)}]{skipetrov2011eigenvalue}%
  \BibitemOpen
  \bibfield  {author} {\bibinfo {author} {\bibfnamefont {S.}~\bibnamefont
  {Skipetrov}}\ and\ \bibinfo {author} {\bibfnamefont {A.}~\bibnamefont
  {Goetschy}},\ }\bibfield  {title} {\emph {\bibinfo {title} {Eigenvalue
  distributions of large Euclidean random matrices for waves in random
  media},}\ }\href@noop {} {\bibfield  {journal} {\bibinfo  {journal} {Journal
  of Physics A: Mathematical and Theoretical}\ }\textbf {\bibinfo {volume}
  {44}},\ \bibinfo {pages} {065102} (\bibinfo {year} {2011})}\BibitemShut
  {NoStop}%
\bibitem [{\citenamefont {M{\'a}ximo}\ \emph {et~al.}(2015)\citenamefont
  {M{\'a}ximo}, \citenamefont {Piovella}, \citenamefont {Courteille},
  \citenamefont {Kaiser},\ and\ \citenamefont {Bachelard}}]{maximo2015spatial}%
  \BibitemOpen
  \bibfield  {author} {\bibinfo {author} {\bibfnamefont {C.~E.}\ \bibnamefont
  {M{\'a}ximo}}, \bibinfo {author} {\bibfnamefont {N.}~\bibnamefont
  {Piovella}}, \bibinfo {author} {\bibfnamefont {P.~W.}\ \bibnamefont
  {Courteille}}, \bibinfo {author} {\bibfnamefont {R.}~\bibnamefont {Kaiser}},
  \ and\ \bibinfo {author} {\bibfnamefont {R.}~\bibnamefont {Bachelard}},\
  }\bibfield  {title} {\emph {\bibinfo {title} {Spatial and temporal
  localization of light in two dimensions},}\ }\href@noop {} {\bibfield
  {journal} {\bibinfo  {journal} {Physical Review A}\ }\textbf {\bibinfo
  {volume} {92}},\ \bibinfo {pages} {062702} (\bibinfo {year}
  {2015})}\BibitemShut {NoStop}%
\bibitem [{\citenamefont {Lawrence}\ \emph
  {et~al.}(2012{\natexlab{b}})\citenamefont {Lawrence}, \citenamefont
  {Trevino},\ and\ \citenamefont {{Dal Negro}}}]{Lawrence2012pillars}%
  \BibitemOpen
  \bibfield  {author} {\bibinfo {author} {\bibfnamefont {N.}~\bibnamefont
  {Lawrence}}, \bibinfo {author} {\bibfnamefont {J.}~\bibnamefont {Trevino}}, \
  and\ \bibinfo {author} {\bibfnamefont {L.}~\bibnamefont {{Dal Negro}}},\
  }\bibfield  {title} {\emph {\bibinfo {title} {{Aperiodic arrays of active
  nanopillars for radiation engineering}},}\ }\href@noop {} {\bibfield
  {journal} {\bibinfo  {journal} {Journal of Applied Physics}\ }\textbf
  {\bibinfo {volume} {111}},\ \bibinfo {pages} {113101} (\bibinfo {year}
  {2012}{\natexlab{b}})}\BibitemShut {NoStop}%
\bibitem [{\citenamefont {Aubry}\ \emph {et~al.}(2020)\citenamefont {Aubry},
  \citenamefont {Froufe-P{\'e}rez}, \citenamefont {Kuhl}, \citenamefont
  {Legrand}, \citenamefont {Scheffold},\ and\ \citenamefont
  {Mortessagne}}]{aubry2020experimental}%
  \BibitemOpen
  \bibfield  {author} {\bibinfo {author} {\bibfnamefont {G.~J.}\ \bibnamefont
  {Aubry}}, \bibinfo {author} {\bibfnamefont {L.~S.}\ \bibnamefont
  {Froufe-P{\'e}rez}}, \bibinfo {author} {\bibfnamefont {U.}~\bibnamefont
  {Kuhl}}, \bibinfo {author} {\bibfnamefont {O.}~\bibnamefont {Legrand}},
  \bibinfo {author} {\bibfnamefont {F.}~\bibnamefont {Scheffold}}, \ and\
  \bibinfo {author} {\bibfnamefont {F.}~\bibnamefont {Mortessagne}},\
  }\bibfield  {title} {\emph {\bibinfo {title} {Experimental tuning of
  transport regimes in hyperuniform disordered photonic materials},}\
  }\href@noop {} {\bibfield  {journal} {\bibinfo  {journal} {Physical Review
  Letters}\ }\textbf {\bibinfo {volume} {125}},\ \bibinfo {pages} {127402}
  (\bibinfo {year} {2020})}\BibitemShut {NoStop}%
\bibitem [{\citenamefont {Busch}\ \emph {et~al.}(2007)\citenamefont {Busch},
  \citenamefont {Von~Freymann}, \citenamefont {Linden}, \citenamefont
  {Mingaleev}, \citenamefont {Tkeshelashvili},\ and\ \citenamefont
  {Wegener}}]{busch2007periodic}%
  \BibitemOpen
  \bibfield  {author} {\bibinfo {author} {\bibfnamefont {K.}~\bibnamefont
  {Busch}}, \bibinfo {author} {\bibfnamefont {G.}~\bibnamefont {Von~Freymann}},
  \bibinfo {author} {\bibfnamefont {S.}~\bibnamefont {Linden}}, \bibinfo
  {author} {\bibfnamefont {S.}~\bibnamefont {Mingaleev}}, \bibinfo {author}
  {\bibfnamefont {L.}~\bibnamefont {Tkeshelashvili}}, \ and\ \bibinfo {author}
  {\bibfnamefont {M.}~\bibnamefont {Wegener}},\ }\bibfield  {title} {\emph
  {\bibinfo {title} {Periodic nanostructures for photonics},}\ }\href@noop {}
  {\bibfield  {journal} {\bibinfo  {journal} {Physics reports}\ }\textbf
  {\bibinfo {volume} {444}},\ \bibinfo {pages} {101} (\bibinfo {year}
  {2007})}\BibitemShut {NoStop}%
\bibitem [{\citenamefont {Asatryan}\ \emph {et~al.}(2001)\citenamefont
  {Asatryan}, \citenamefont {Busch}, \citenamefont {McPhedran}, \citenamefont
  {Botten}, \citenamefont {de~Sterke},\ and\ \citenamefont
  {Nicorovici}}]{AsatryanPRE}%
  \BibitemOpen
  \bibfield  {author} {\bibinfo {author} {\bibfnamefont {A.}~\bibnamefont
  {Asatryan}}, \bibinfo {author} {\bibfnamefont {K.}~\bibnamefont {Busch}},
  \bibinfo {author} {\bibfnamefont {R.}~\bibnamefont {McPhedran}}, \bibinfo
  {author} {\bibfnamefont {L.}~\bibnamefont {Botten}}, \bibinfo {author}
  {\bibfnamefont {C.~M.}\ \bibnamefont {de~Sterke}}, \ and\ \bibinfo {author}
  {\bibfnamefont {N.}~\bibnamefont {Nicorovici}},\ }\bibfield  {title} {\emph
  {\bibinfo {title} {Two-dimensional Green's function and local density of
  states in photonic crystals consisting of a finite number of cylinders of
  infinite length},}\ }\href@noop {} {\bibfield  {journal} {\bibinfo  {journal}
  {Physical Review E}\ }\textbf {\bibinfo {volume} {63}},\ \bibinfo {pages}
  {046612} (\bibinfo {year} {2001})}\BibitemShut {NoStop}%
\bibitem [{\citenamefont {Rusek}\ and\ \citenamefont
  {Or{\l}owski}(1995)}]{RusekPRE2D}%
  \BibitemOpen
  \bibfield  {author} {\bibinfo {author} {\bibfnamefont {M.}~\bibnamefont
  {Rusek}}\ and\ \bibinfo {author} {\bibfnamefont {A.}~\bibnamefont
  {Or{\l}owski}},\ }\bibfield  {title} {\emph {\bibinfo {title} {Analytical
  approach to localization of electromagnetic waves in two-dimensional random
  media},}\ }\href@noop {} {\bibfield  {journal} {\bibinfo  {journal} {Physical
  Review E}\ }\textbf {\bibinfo {volume} {51}},\ \bibinfo {pages} {R2763}
  (\bibinfo {year} {1995})}\BibitemShut {NoStop}%
\bibitem [{\citenamefont {Skipetrov}\ and\ \citenamefont
  {Sokolov}(2014)}]{SkipetrovPRL}%
  \BibitemOpen
  \bibfield  {author} {\bibinfo {author} {\bibfnamefont {S.~E.}\ \bibnamefont
  {Skipetrov}}\ and\ \bibinfo {author} {\bibfnamefont {I.~M.}\ \bibnamefont
  {Sokolov}},\ }\bibfield  {title} {\emph {\bibinfo {title} {Absence of
  Anderson localization of light in a random ensemble of point scatterers},}\
  }\href@noop {} {\bibfield  {journal} {\bibinfo  {journal} {Physical Review
  Letters}\ }\textbf {\bibinfo {volume} {112}},\ \bibinfo {pages} {023905}
  (\bibinfo {year} {2014})}\BibitemShut {NoStop}%
\bibitem [{\citenamefont {Leseur}\ \emph {et~al.}(2016)\citenamefont {Leseur},
  \citenamefont {Pierrat},\ and\ \citenamefont {Carminati}}]{Leseur}%
  \BibitemOpen
  \bibfield  {author} {\bibinfo {author} {\bibfnamefont {O.}~\bibnamefont
  {Leseur}}, \bibinfo {author} {\bibfnamefont {R.}~\bibnamefont {Pierrat}}, \
  and\ \bibinfo {author} {\bibfnamefont {R.}~\bibnamefont {Carminati}},\
  }\bibfield  {title} {\emph {\bibinfo {title} {High-density hyperuniform
  materials can be transparent},}\ }\href@noop {} {\bibfield  {journal}
  {\bibinfo  {journal} {Optica}\ }\textbf {\bibinfo {volume} {3}},\ \bibinfo
  {pages} {763} (\bibinfo {year} {2016})}\BibitemShut {NoStop}%
\bibitem [{\citenamefont {Caz{\'e}}\ \emph {et~al.}(2013)\citenamefont
  {Caz{\'e}}, \citenamefont {Pierrat},\ and\ \citenamefont {Carminati}}]{Caze}%
  \BibitemOpen
  \bibfield  {author} {\bibinfo {author} {\bibfnamefont {A.}~\bibnamefont
  {Caz{\'e}}}, \bibinfo {author} {\bibfnamefont {R.}~\bibnamefont {Pierrat}}, \
  and\ \bibinfo {author} {\bibfnamefont {R.}~\bibnamefont {Carminati}},\
  }\bibfield  {title} {\emph {\bibinfo {title} {Strong coupling to
  two-dimensional Anderson localized modes},}\ }\href@noop {} {\bibfield
  {journal} {\bibinfo  {journal} {Physical Review Letters}\ }\textbf {\bibinfo
  {volume} {111}},\ \bibinfo {pages} {053901} (\bibinfo {year}
  {2013})}\BibitemShut {NoStop}%
\bibitem [{\citenamefont {Bouchet}\ and\ \citenamefont
  {Carminati}(2019)}]{Bouchet}%
  \BibitemOpen
  \bibfield  {author} {\bibinfo {author} {\bibfnamefont {D.}~\bibnamefont
  {Bouchet}}\ and\ \bibinfo {author} {\bibfnamefont {R.}~\bibnamefont
  {Carminati}},\ }\bibfield  {title} {\emph {\bibinfo {title} {Quantum dipole
  emitters in structured environments: a scattering approach: tutorial},}\
  }\href@noop {} {\bibfield  {journal} {\bibinfo  {journal} {Journal of the
  Optical Society of America B}\ }\textbf {\bibinfo {volume} {36}},\ \bibinfo
  {pages} {186} (\bibinfo {year} {2019})}\BibitemShut {NoStop}%
\bibitem [{\citenamefont {Lagendijk}\ and\ \citenamefont
  {Van~Tiggelen}(1996)}]{Lagendijk}%
  \BibitemOpen
  \bibfield  {author} {\bibinfo {author} {\bibfnamefont {A.}~\bibnamefont
  {Lagendijk}}\ and\ \bibinfo {author} {\bibfnamefont {B.~A.}\ \bibnamefont
  {Van~Tiggelen}},\ }\bibfield  {title} {\emph {\bibinfo {title} {Resonant
  multiple scattering of light},}\ }\href@noop {} {\bibfield  {journal}
  {\bibinfo  {journal} {Physics Reports}\ }\textbf {\bibinfo {volume} {270}},\
  \bibinfo {pages} {143} (\bibinfo {year} {1996})}\BibitemShut {NoStop}%
\bibitem [{\citenamefont {Gagnon}\ and\ \citenamefont
  {Dub{\'e}}(2015)}]{Gagnon}%
  \BibitemOpen
  \bibfield  {author} {\bibinfo {author} {\bibfnamefont {D.}~\bibnamefont
  {Gagnon}}\ and\ \bibinfo {author} {\bibfnamefont {L.~J.}\ \bibnamefont
  {Dub{\'e}}},\ }\bibfield  {title} {\emph {\bibinfo {title} {Lorenz--Mie
  theory for 2D scattering and resonance calculations},}\ }\href@noop {}
  {\bibfield  {journal} {\bibinfo  {journal} {Journal of Optics}\ }\textbf
  {\bibinfo {volume} {17}},\ \bibinfo {pages} {103501} (\bibinfo {year}
  {2015})}\BibitemShut {NoStop}%
\bibitem [{\citenamefont {Doicu}\ \emph {et~al.}(2006)\citenamefont {Doicu},
  \citenamefont {Wriedt},\ and\ \citenamefont {Eremin}}]{doicu2006light}%
  \BibitemOpen
  \bibfield  {author} {\bibinfo {author} {\bibfnamefont {A.}~\bibnamefont
  {Doicu}}, \bibinfo {author} {\bibfnamefont {T.}~\bibnamefont {Wriedt}}, \
  and\ \bibinfo {author} {\bibfnamefont {Y.~A.}\ \bibnamefont {Eremin}},\
  }\href@noop {} {\emph {\bibinfo {title} {Light scattering by systems of
  particles: null-field method with discrete sources: theory and programs}}},\
  Vol.\ \bibinfo {volume} {124}\ (\bibinfo  {publisher} {Springer},\ \bibinfo
  {year} {2006})\BibitemShut {NoStop}%
\bibitem [{\citenamefont {Wriedt}(2012)}]{Wriedt2012}%
  \BibitemOpen
  \bibfield  {author} {\bibinfo {author} {\bibfnamefont {T.}~\bibnamefont
  {Wriedt}},\ }\emph {\bibinfo {title} {Mie Theory: A Review},}\ in\ \href@noop
  {} {\emph {\bibinfo {booktitle} {The Mie Theory: Basics and Applications}}},\
  \bibinfo {editor} {edited by\ \bibinfo {editor} {\bibfnamefont
  {W.}~\bibnamefont {Hergert}}\ and\ \bibinfo {editor} {\bibfnamefont
  {T.}~\bibnamefont {Wriedt}}}\ (\bibinfo  {publisher} {Springer},\ \bibinfo
  {year} {2012})\ pp.\ \bibinfo {pages} {53--71}\BibitemShut {NoStop}%
\bibitem [{\citenamefont {Trojak}\ \emph {et~al.}(2021)\citenamefont {Trojak},
  \citenamefont {Gorsky}, \citenamefont {Murray}, \citenamefont {Sgrignuoli},
  \citenamefont {Pinheiro}, \citenamefont {Dal~Negro},\ and\ \citenamefont
  {Sapienza}}]{Trojak1}%
  \BibitemOpen
  \bibfield  {author} {\bibinfo {author} {\bibfnamefont {O.~J.}\ \bibnamefont
  {Trojak}}, \bibinfo {author} {\bibfnamefont {S.}~\bibnamefont {Gorsky}},
  \bibinfo {author} {\bibfnamefont {C.}~\bibnamefont {Murray}}, \bibinfo
  {author} {\bibfnamefont {F.}~\bibnamefont {Sgrignuoli}}, \bibinfo {author}
  {\bibfnamefont {F.~A.}\ \bibnamefont {Pinheiro}}, \bibinfo {author}
  {\bibfnamefont {L.}~\bibnamefont {Dal~Negro}}, \ and\ \bibinfo {author}
  {\bibfnamefont {L.}~\bibnamefont {Sapienza}},\ }\bibfield  {title} {\emph
  {\bibinfo {title} {Cavity-enhanced light--matter interaction in Vogel-spiral
  devices as a platform for quantum photonics},}\ }\href@noop {} {\bibfield
  {journal} {\bibinfo  {journal} {Applied Physics Letters}\ }\textbf {\bibinfo
  {volume} {118}},\ \bibinfo {pages} {011103} (\bibinfo {year}
  {2021})}\BibitemShut {NoStop}%
\bibitem [{\citenamefont {Trojak}\ \emph {et~al.}(2020)\citenamefont {Trojak},
  \citenamefont {Gorsky}, \citenamefont {Sgrignuoli}, \citenamefont {Pinheiro},
  \citenamefont {Park}, \citenamefont {Song}, \citenamefont {Dal~Negro},\ and\
  \citenamefont {Sapienza}}]{Trojak2}%
  \BibitemOpen
  \bibfield  {author} {\bibinfo {author} {\bibfnamefont {O.~J.}\ \bibnamefont
  {Trojak}}, \bibinfo {author} {\bibfnamefont {S.}~\bibnamefont {Gorsky}},
  \bibinfo {author} {\bibfnamefont {F.}~\bibnamefont {Sgrignuoli}}, \bibinfo
  {author} {\bibfnamefont {F.~A.}\ \bibnamefont {Pinheiro}}, \bibinfo {author}
  {\bibfnamefont {S.-I.}\ \bibnamefont {Park}}, \bibinfo {author}
  {\bibfnamefont {J.~D.}\ \bibnamefont {Song}}, \bibinfo {author}
  {\bibfnamefont {L.}~\bibnamefont {Dal~Negro}}, \ and\ \bibinfo {author}
  {\bibfnamefont {L.}~\bibnamefont {Sapienza}},\ }\bibfield  {title} {\emph
  {\bibinfo {title} {Cavity quantum electro-dynamics with solid-state emitters
  in aperiodic nano-photonic spiral devices},}\ }\href@noop {} {\bibfield
  {journal} {\bibinfo  {journal} {Applied Physics Letters}\ }\textbf {\bibinfo
  {volume} {117}},\ \bibinfo {pages} {124006} (\bibinfo {year}
  {2020})}\BibitemShut {NoStop}%
\bibitem [{\citenamefont {Ringler}\ \emph {et~al.}(2008)\citenamefont
  {Ringler}, \citenamefont {Schwemer}, \citenamefont {Wunderlich},
  \citenamefont {Nichtl}, \citenamefont {K{\"u}rzinger}, \citenamefont {Klar},\
  and\ \citenamefont {Feldmann}}]{ringler2008shaping}%
  \BibitemOpen
  \bibfield  {author} {\bibinfo {author} {\bibfnamefont {M.}~\bibnamefont
  {Ringler}}, \bibinfo {author} {\bibfnamefont {A.}~\bibnamefont {Schwemer}},
  \bibinfo {author} {\bibfnamefont {M.}~\bibnamefont {Wunderlich}}, \bibinfo
  {author} {\bibfnamefont {A.}~\bibnamefont {Nichtl}}, \bibinfo {author}
  {\bibfnamefont {K.}~\bibnamefont {K{\"u}rzinger}}, \bibinfo {author}
  {\bibfnamefont {T.}~\bibnamefont {Klar}}, \ and\ \bibinfo {author}
  {\bibfnamefont {J.}~\bibnamefont {Feldmann}},\ }\bibfield  {title} {\emph
  {\bibinfo {title} {Shaping emission spectra of fluorescent molecules with
  single plasmonic nanoresonators},}\ }\href@noop {} {\bibfield  {journal}
  {\bibinfo  {journal} {Physical Review Letters}\ }\textbf {\bibinfo {volume}
  {100}},\ \bibinfo {pages} {203002} (\bibinfo {year} {2008})}\BibitemShut
  {NoStop}%
\bibitem [{\citenamefont {Vogel}\ and\ \citenamefont
  {Welsch}(2006)}]{vogel2006quantum}%
  \BibitemOpen
  \bibfield  {author} {\bibinfo {author} {\bibfnamefont {W.}~\bibnamefont
  {Vogel}}\ and\ \bibinfo {author} {\bibfnamefont {D.-G.}\ \bibnamefont
  {Welsch}},\ }\href@noop {} {\emph {\bibinfo {title} {Quantum optics}}}\
  (\bibinfo  {publisher} {John Wiley \& Sons},\ \bibinfo {year}
  {2006})\BibitemShut {NoStop}%
\bibitem [{\citenamefont {Sgrignuoli}\ \emph {et~al.}(2015)\citenamefont
  {Sgrignuoli}, \citenamefont {Mazzamuto}, \citenamefont {Caselli},
  \citenamefont {Intonti}, \citenamefont {Cataliotti}, \citenamefont
  {Gurioli},\ and\ \citenamefont {Toninelli}}]{SgrignuoliACS}%
  \BibitemOpen
  \bibfield  {author} {\bibinfo {author} {\bibfnamefont {F.}~\bibnamefont
  {Sgrignuoli}}, \bibinfo {author} {\bibfnamefont {G.}~\bibnamefont
  {Mazzamuto}}, \bibinfo {author} {\bibfnamefont {N.}~\bibnamefont {Caselli}},
  \bibinfo {author} {\bibfnamefont {F.}~\bibnamefont {Intonti}}, \bibinfo
  {author} {\bibfnamefont {F.~S.}\ \bibnamefont {Cataliotti}}, \bibinfo
  {author} {\bibfnamefont {M.}~\bibnamefont {Gurioli}}, \ and\ \bibinfo
  {author} {\bibfnamefont {C.}~\bibnamefont {Toninelli}},\ }\bibfield  {title}
  {\emph {\bibinfo {title} {Necklace state hallmark in disordered 2D photonic
  systems},}\ }\href@noop {} {\bibfield  {journal} {\bibinfo  {journal} {ACS
  Photonics}\ }\textbf {\bibinfo {volume} {2}},\ \bibinfo {pages} {1636}
  (\bibinfo {year} {2015})}\BibitemShut {NoStop}%
\bibitem [{\citenamefont {Mirlin}(2000)}]{Mirlin}%
  \BibitemOpen
  \bibfield  {author} {\bibinfo {author} {\bibfnamefont {A.~D.}\ \bibnamefont
  {Mirlin}},\ }\bibfield  {title} {\emph {\bibinfo {title} {Statistics of
  energy levels and eigenfunctions in disordered systems},}\ }\href@noop {}
  {\bibfield  {journal} {\bibinfo  {journal} {Physics Reports}\ }\textbf
  {\bibinfo {volume} {326}},\ \bibinfo {pages} {259} (\bibinfo {year}
  {2000})}\BibitemShut {NoStop}%
\bibitem [{\citenamefont {Skipetrov}(2016)}]{Skipetrov2016}%
  \BibitemOpen
  \bibfield  {author} {\bibinfo {author} {\bibfnamefont {S.~E.}\ \bibnamefont
  {Skipetrov}},\ }\bibfield  {title} {\emph {\bibinfo {title} {{Finite-size
  scaling analysis of localization transition for scalar waves in a
  three-dimensional ensemble of resonant point scatterers}},}\ }\href {\doibase
  10.1103/PhysRevB.94.064202} {\bibfield  {journal} {\bibinfo  {journal}
  {Physical Review B}\ }\textbf {\bibinfo {volume} {94}},\ \bibinfo {pages}
  {064202} (\bibinfo {year} {2016})}\BibitemShut {NoStop}%
\bibitem [{\citenamefont {Trevino}\ \emph {et~al.}(2012)\citenamefont
  {Trevino}, \citenamefont {Liew}, \citenamefont {Noh}, \citenamefont {Cao},\
  and\ \citenamefont {Dal~Negro}}]{TrevinoMF}%
  \BibitemOpen
  \bibfield  {author} {\bibinfo {author} {\bibfnamefont {J.}~\bibnamefont
  {Trevino}}, \bibinfo {author} {\bibfnamefont {S.~F.}\ \bibnamefont {Liew}},
  \bibinfo {author} {\bibfnamefont {H.}~\bibnamefont {Noh}}, \bibinfo {author}
  {\bibfnamefont {H.}~\bibnamefont {Cao}}, \ and\ \bibinfo {author}
  {\bibfnamefont {L.}~\bibnamefont {Dal~Negro}},\ }\bibfield  {title} {\emph
  {\bibinfo {title} {Geometrical structure, multifractal spectra and localized
  optical modes of aperiodic Vogel spirals},}\ }\href@noop {} {\bibfield
  {journal} {\bibinfo  {journal} {Optics express}\ }\textbf {\bibinfo {volume}
  {20}},\ \bibinfo {pages} {3015} (\bibinfo {year} {2012})}\BibitemShut
  {NoStop}%
\bibitem [{\citenamefont {Varga}\ and\ \citenamefont
  {Pipek}(2003)}]{VargaRenyi}%
  \BibitemOpen
  \bibfield  {author} {\bibinfo {author} {\bibfnamefont {I.}~\bibnamefont
  {Varga}}\ and\ \bibinfo {author} {\bibfnamefont {J.}~\bibnamefont {Pipek}},\
  }\bibfield  {title} {\emph {\bibinfo {title} {R{\'e}nyi entropies
  characterizing the shape and the extension of the phase space representation
  of quantum wave functions in disordered systems},}\ }\href@noop {} {\bibfield
   {journal} {\bibinfo  {journal} {Physical Review E}\ }\textbf {\bibinfo
  {volume} {68}},\ \bibinfo {pages} {026202} (\bibinfo {year}
  {2003})}\BibitemShut {NoStop}%
\end{thebibliography}
